\documentclass[superscriptaddress,showpacs,twocolumn,preprintnumbers]{revtex4}

\usepackage{amsmath}

\usepackage{float}
\usepackage{graphicx}	
\usepackage{hyperref}
\usepackage{mathtools}
\usepackage{cancel}
\usepackage{multirow}
\usepackage{rotating}
\usepackage{physics}

\begin{document}

\title{Exact Oscillation Probabilities of Neutrinos in Three generations \\ derived from Relativistic Equation}

\author{Keiichi Kimura}
\email{kimukei@eken.phys.nagoya-u.ac.jp}
\affiliation{Department of Physics, Nagoya University, Nagoya, 464-8602,
Japan}

\author{Akira Takamura}
\email{takamura@eken.phys.nagoya-u.ac.jp}
\affiliation{Department of Physics, Nagoya University, Nagoya, 464-8602,
Japan}
\affiliation{Department of Mathematics,
Toyota National College of Technology, Eisei-cho 2-1, Toyota-shi, 471-8525, Japan}

\date{\today}
\begin{abstract}
In three generations or more, we derive the oscillation probabilities of both Dirac and Majorana neutrinos relativistically by using the Dirac equation.
We present various oscillation probabilities for including wrong-helicity neutrinos, right-handed neutrinos, and anti-neutrinos.
We summarize the relations between these probabilities.
As neutrinos have finite mass, there are two components for each chirality
corresponding to positive and negative helicities.
We show that the probability is different for each component even if neutrinos have the same chirality.
The probabilities derived by the relativistic equation depend on not only the mass squared differences but also the absolute masses of neutrinos.
Besides, the new CP phases appear in the probabilities of oscillations with chirality-flip.
These new CP phases are equivalent to the Majorana CP phases in the case of Majorana neutrinos.
We investigate the CP dependence of oscillation probabilities in vacuum.
There are no direct CP violation in $\nu_{\alpha}\leftrightarrow \nu_{\beta}^c$ oscillations even if the flavors, $\alpha$ and $\beta$, are different as in the same as two generations.
In other words, the difference between the CP-conjugate probabilities vanishes.
However, in three generations or more, the sine terms of new CP phases appear in the probabilities in addition to the cosine terms. 
This is different from the result obtained in two generations.
Furthermore, the zero-distance effect does not appear in our formulation.
\end{abstract} 

\maketitle

\section{Introduction}
\label{sec:introduction } 

The idea of neutrino-antineutrino oscillations proposed by Pontecorvo in 1957 \cite{Pontecorvo}. After the discovery of muon neutrino, Maki, Nakagawa and Sakata \cite{MNS} proposed the oscillations between neutrinos with different flavors in 1962, and the oscillations have been confirmed in the Super-Kamiokande atmospheric neutrino experiment in 1998 \cite{1998SK}.
In the period of twenty years after the experiment, the evidence of neutrino oscillations has been accumulating
in the solar neutrino experiments \cite{SK, SNO, SK2}, the long-baseline experiments \cite{T2K, MINOS},
and the reactor experiments \cite{KamLAND, DayaBay, RENO, DoubleChooz}.
The understanding of the neutrino mass squared differences and mixing angles has proceeded
through these experiments
and we are getting the clue of the leptonic Dirac CP phase at present \cite{Dirac CP, NOvA}.
To estimate the value of the Dirac CP phase as precisely as possible,
the exact formulation of the oscillation probabilities including matter effect
has been developed \cite{Zaglauer, Ohlsson, KTY, Yokomakura0207, Yasuda}.

On the other hand, 0$\nu\beta\beta$ decay experiments have been performed
to determine whether the neutrino is the Dirac particle or the Majorana particle \cite{Majorana}
and the absolute value of neutrino mass \cite{KamLAND-Zen:2016pfg, Alfonso:2015wka, Albert:2014awa, Agostini:2013mzu, Gando:2012zm,
Elliott:2016ble, Andringa:2015tza}.
The possibility for the transition from neutrinos to anti-neutrinos with different flavor
were also discussed to investigate the Majorana CP phases
\cite{Bahcall1978, Valle1981, Li1982, Bernabeu1983, Gouvea2003, Xing2013}.

In our previous papers,
we have derived the exact neutrino oscillation probabilities relativistically by using the Dirac equation
to analyze future neutrino experiments as precisely as possible.
In the first paper, we gave the formulation for the Dirac neutrinos in two generations \cite{KT1}.
As the result, a new CP phase different from the Dirac CP phase appears in the oscillations with chirality-flip
even in the framework of two-generation Dirac neutrinos.
We have also shown that the terms dependent on the absolute value of neutrino mass also appear
in our formulation.
In the second paper, we applied the relativistic formulation for the Majorana neutrinos \cite{KT2}.
We have shown the new phase that appeared in the case of the Dirac neutrinos becomes the Majorana CP phase.
This is because $\nu_L^c$ in Majorana neutrinos plays a role of $\nu_R$ in Dirac neutrinos.
We can interpret that the Majorana CP phase is not accompanied by the lepton number violation
but with the chirality-flip.

In this paper, we extend our relativistic formulation to three generations or more.
We derive various oscillation probabilities for including wrong-helicity neutrinos, right-handed neutrinos, and anti-neutrinos in a unified way.
We summarize the relations in these probabilities.
In the Dirac equation, there are two components for each chirality and each generation corresponding to positive and negative helicities.
We show that the probabilities with different helicities are not the same even if the neutrinos have the same chirality and the same flavor.
The probabilities derived by the relativistic equation depend on not only the mass squared differences but also the absolute masses of neutrinos.
Besides, the new CP phases appear in the probabilities of oscillations with chirality-flip.
These new CP phases are equivalent to the Majorana CP phases in the case of Majorana neutrinos.
We investigate the CP dependence of oscillation probabilities in vacuum.
In the case of Majorana neutrinos, there is no direct CP violation in $\nu_{\alpha}\leftrightarrow \nu_{\beta}^c$ oscillations even if the flavors, $\alpha$ and $\beta$, are different as in the same as two generations \cite{KT2}.
In other words, the difference between the CP-conjugate probabilities, 
$P(\nu_{\alpha L}\to \nu_{\beta L}^c)-P(\nu_{\alpha L}^c\to \nu_{\beta L})$ vanishes.
However, in three generations or more, the sine terms of new CP phases appear in the probabilities in addition to the cosine terms.
This is different from the result obtained in two generations.
Furthermore, the zero-distance effect \cite{Li1982}, which was known as the phenomena for neutrinos
instantly changing to anti-neutrinos, cannot be occured from our calculation.
These results are different from the previous ones.

We give the number of the independent CP phases in n-generations.
For both Dirac and Majorana neutrinos, the number of the Dirac CP phases is given by
\begin{eqnarray}
\frac{(n-1)(n-2)}{2},
\end{eqnarray}
and the number of the CP phases accompanied to the oscillations with chirality-flip is
\begin{eqnarray}
n-1.
\end{eqnarray}
Therefore, the total number of the independent CP phases becomes
\begin{eqnarray}
\frac{n(n-1)}{2},
\end{eqnarray}
and in accordance with the result of the Majorana neutrinos \cite{Majorana-phase}.
If neutrinos are the Dirac particles and the flavors of $\nu_R$ cannot be distinguished
beyond the Standard Model,
$(n-1)$ CP phases originated from the oscillations with the chirality-flip are not observable
and coincide with the previous result.

The paper is organized as follows. In section II, we define our notations used in this paper.
In section III, we review the non-relativistic derivation of neutrino oscillation probabilities
developed in the previous papers by using the Schr${\ddot{\rm o}}$dinger equation. In section IV,
we present the relativistic derivation of various neutrino oscillation probabilities for Dirac neutrinos
including wrong-helicity neutrinos, right-handed neutrinos and anti-neutrinos by using the Dirac equation.
We also investigate the CP dependence of the probabilities, in particular on the new CP phases, and count the number of independent CP phases.
In section V, we also present the relativistic derivation of oscillation probabilities
for Majorana neutrinos.
In section VI, we summarize the relation of these oscillation probabilities.
In section VII, we compare our result of the Majorana neutrinos with the previous one.
In section VIII, we summarize our results obtained in this paper.

%%%%%%%%%%%%%%%%%%%%%%%%%%%%%%%%%%%%%%%%%%%%%%%%%%%
\section{Notation}

In this section, we write down the notation used in this paper. 
We mainly use the chiral representation 
because neutrinos are measured through weak interaction.
In chiral representation, the gamma matrices with $4\times 4$ form are given by 
\begin{eqnarray}
\gamma^0=\left(\begin{array}{cc}0 & 1 \\ 1 & 0\end{array}\right), \,
\gamma^i=\left(\begin{array}{cc}0 & -\sigma_i \\ \sigma_i & 0\end{array}\right), \, 
\gamma_5=\left(\begin{array}{cc}1 & 0 \\ 0 & -1\end{array}\right),  \label{gamma-mat}
\end{eqnarray}
where $2\times 2$ $\sigma_i$ matrices are defined by 
\begin{eqnarray}
\sigma_1=\left(\begin{array}{cc}0 & 1 \\ 1 & 0\end{array}\right), \,
\sigma_2=\left(\begin{array}{cc}0 & -i \\ i & 0\end{array}\right), \, 
\sigma_3=\left(\begin{array}{cc}1 & 0 \\ 0 & -1\end{array}\right). 
\end{eqnarray}
We also define 4-component spinors $\psi$, $\psi_L$ and $\psi_R$ as 
\begin{eqnarray}
&&\hspace{-0.5cm}\psi=\left(\begin{array}{c}\xi \\ \eta \end{array}\right), \\
&&\hspace{-0.5cm}\psi_L=\frac{1-\gamma_5}{2}\psi=\left(\begin{array}{c}0 \\ \eta\end{array}\right), \, 
\psi_R=\frac{1+\gamma_5}{2}\psi=\left(\begin{array}{c}\xi \\ 0 \end{array}\right),   \label{psi-def}
\end{eqnarray}
and 2-component spinors $\xi$ and $\eta$ as 
\begin{eqnarray}
\xi=\left(\begin{array}{c}\nu_R^{\prime} \\ \nu_R \end{array}\right), \qquad 
\eta=\left(\begin{array}{c}\nu_L^{\prime} \\ \nu_L \end{array}\right).  
\end{eqnarray}
Furthermore, 
we use the subscript $\alpha$ and $\beta$ for flavor, $L$ and $R$ for chirality, 
the number $j$ and $k$ for generation and superscript $\pm$ for energy. 
Because of negligible neutrino mass, mass eigenstate has been often identified with energy eigenstate 
in the previous papers.
But in the future, we should distinguish these two kinds of eigenstates for the finite neutrino mass. 
More concretely, we use the following eigenstates; 
\begin{eqnarray}
&&{\rm chirality\mathchar`-flavor \,\, eigenstates}:  \nu_{\alpha L}, \nu_{\alpha R}, \nu_{\beta L}, \nu_{\beta R}, \\
&&{\rm chirality\mathchar`-mass \,\, eigenstates}: \,\, \nu_{jL}, \nu_{jR}, \nu_{kL}, \nu_{kR}, \\
&&{\rm energy\mathchar`-helicity \,\, eigenstates}: \nu_j^+, \nu_j^-, \nu_k^+, \nu_k^-.
\end{eqnarray}
It is noted that chirality-mass eigenstates are not exactly the eigenstates of the Hamiltonian.
We use the term, eigenstates, in the sense that the mass submatrix in the Hamiltonian is diagonalized.
Judging from common sense, one may think it strange that the chirality and the mass live in the same eigenstate. 
Details will be explained in the subsequent section.

We also difine the spinor for anti-neutrino as charge conjugation of neutrino $\psi^c=i\gamma^2 \psi^*$.
The charge conjugations for left-handed and right-handed neutrinos are defined by 
\begin{eqnarray}
&&\hspace{-0.5cm}\psi_L^c\equiv(\psi_L)^c\equiv
\left(\begin{array}{c}\nu_L^c \\ \nu_L^{c\prime} \\ 0 \\ 0 \end{array}\right) 
\equiv i\gamma^2 \psi_L^*=i\gamma^2 \frac{1-\gamma_5}{2}\psi^*  \nonumber \\
&&\hspace{-0.5cm}=\frac{1+\gamma_5}{2}(i\gamma^2 \psi^*) 
=(\psi^c)_R=\left(\!\!\!\begin{array}{c}i\sigma_2 \eta^* \\ 0 \end{array}\!\!\!\right)
=\left(\!\!\begin{array}{c}\nu_L^* \\ -\nu_L^{*\prime} \\ 0 \\ 0 \end{array}\!\!\right), \label{nuc} \\
&&\hspace{-0.5cm}\psi_R^c\equiv(\psi_R)^c\equiv\left(\begin{array}{c}0 \\ 0 \\ \nu_R^c \\ \nu_R^{c\prime} \end{array}\right) 
\equiv i\gamma^2 \psi_R^*=i\gamma^2 \frac{1+\gamma_5}{2}\psi^*  \nonumber \\
&&\hspace{-0.5cm}\!=\!\frac{1-\gamma_5}{2}(i\gamma^2 \psi^*) 
\!=\!(\psi^c)_L=\left(\!\!\!\begin{array}{c}0 \\ -i\sigma_2 \xi^* \end{array}\!\!\!\right)
\!=\!\left(\!\!\begin{array}{c}0 \\ 0 \\ -\nu_R^* \\ \nu_R^{*\prime} \end{array}\!\!\right). 
\end{eqnarray}
It is noted that the chirality is flipped by taking the charge conjugation.

%%%%%%%%%%%%%%%%%%%%%%%%%%%%%%%%%%%%%%%%%%%%%%%%%%%%%%%%%%
\section{Review of Oscillation Probabilities from Non-Relativistic Equation}

In this section, we review how the neutrino oscillation probabilities in vacuum were derived in the previous 
papers. 
For example, in ref. \cite{PDG}, the flavor eigenstates are given as the linear combination of 
the energy (mass) eigenstates, 
\begin{eqnarray}
\left(\begin{array}{c}
\nu_{eL} \\ \nu_{\mu L} \\ \nu_{\tau L}
\end{array}\right)
=\left(\begin{array}{ccc}
U_{e1} & U_{e2} & U_{e3} \\ 
U_{\mu 1} & U_{\mu 2} & U_{\mu 3} \\
U_{\tau 1} & U_{\tau 2} & U_{\tau 3}
\end{array}\right)\left(\begin{array}{c}
\nu_{1}^+ \\ \nu_{2}^+ \\ \nu_3^+
\end{array}\right).
\end{eqnarray}
The energy eigenstates evolve following the equation 
\begin{eqnarray}
\frac{d}{dt}\left(\begin{array}{c}
\nu_{1}^+ \\ \nu_{2}^+ \\ \nu_3^+
\end{array}\right)
=\left(\begin{array}{ccc}
E_1 & 0 & 0 \\ 
0 & E_2 & 0 \\
0 & 0 & E_3
\end{array}\right)\left(\begin{array}{c}
\nu_{1}^+ \\ \nu_{2}^+ \\ \nu_3^+
\end{array}\right), 
\end{eqnarray}
and after the time $t$, the flavor eigenstates become 
\begin{widetext}
\begin{eqnarray}
\left(\begin{array}{c}
\nu_{eL}(t) \\ \nu_{\mu L}(t) \\ \nu_{\tau L}(t)
\end{array}\right)
=\left(\begin{array}{ccc}
U_{e1} & U_{e2} & U_{e3} \\ 
U_{\mu 1} & U_{\mu 2} & U_{\mu 3} \\
U_{\tau 1} & U_{\tau 2} & U_{\tau 3}
\end{array}\right)\left(\begin{array}{ccc}
e^{-iE_1 t} & 0 & 0 \\ 
0 & e^{-iE_2 t} & 0 \\
0 & 0 & e^{-iE_3 t}
\end{array}\right)\left(\begin{array}{c}
\nu_{1}^+ \\ \nu_{2}^+ \\ \nu_3^+
\end{array}\right).
\end{eqnarray}
\end{widetext}
Rewriting the relation about the fields to one particle states by using the production operator, we obtain 
\begin{eqnarray}
|\nu_{\alpha L}(t)\rangle=\sum_{j=1}^3 U_{\alpha j}^*e^{-iE_jt}|\nu_j^+\rangle
\end{eqnarray}
and also their conjugate states, 
\begin{eqnarray}
\langle \nu_{\beta L}| =\sum_{j=1}^3U_{\beta j}\langle \nu_j^+|.
\end{eqnarray}
If we take a certain flavor $e$, $\mu$ or $\tau$ as $\alpha$ and $\beta$, the amplitude for $\nu_{\alpha}$ 
to $\nu_{\beta}$ is given by  
\begin{eqnarray}
\hspace{-0.5cm}A(\nu_{\alpha L}\to\nu_{\beta L})=\langle \nu_{\beta L}|\nu_{\alpha L}(t)\rangle 
=\sum_{j=1}^3 U_{\alpha j}^*U_{\beta j}e^{-iE_jt}
\end{eqnarray}
The oscillation probability for $\nu_{\alpha L}$ to $\nu_{\beta L}$ becomes
\begin{widetext}
\begin{eqnarray}
&&P(\nu_{\alpha L}\to\nu_{\beta L})=\left|A(\nu_{\alpha L}\to\nu_{\beta L})\right|^2
=\sum_j |U_{\alpha j}^*U_{\beta j}|^2
+\sum_{j<k}2{\rm Re}[U_{\alpha j}^*U_{\beta j}U_{\alpha k}U_{\beta k}^*e^{-i(E_j-E_k)t}] \nonumber \\
&&\hspace{1cm}=
\sum_j |U_{\alpha j}^*U_{\beta j}|^2
+\sum_{j<k}2\{{\rm Re}[U_{\alpha j}^*U_{\beta j}U_{\alpha k}U_{\beta k}^*]\cos \Delta E_{jk}t
+{\rm Im}[U_{\alpha j}^*U_{\beta j}U_{\alpha k}U_{\beta k}^*]\sin \Delta E_{jk}t\} \nonumber \\
&&\hspace{1cm}=
\left|\sum_j U_{\alpha j}^*U_{\beta j}\right|^2 \!\!
-2\sum_{j<k}{\rm Re}[U_{\alpha j}^*U_{\beta j}U_{\alpha k}U_{\beta k}^*](1-\cos \Delta E_{jk}t)
+2\sum_{j<k}{\rm Im}[U_{\alpha j}^*U_{\beta j}U_{\alpha k}U_{\beta k}^*]\sin \Delta E_{jk}t \nonumber \\
&&\hspace{1cm}=
\delta_{\alpha\beta}-4\sum_{j<k}{\rm Re}[U_{\alpha j}U_{\beta j}^*U_{\alpha k}^*U_{\beta k}]\sin^2 \left(\frac{\Delta E_{jk}t}{2}\right)
-2\sum_{j<k}{\rm Im}[U_{\alpha j}U_{\beta j}^*U_{\alpha k}^*U_{\beta k}]\sin \Delta E_{jk}t,
\end{eqnarray}
where $\Delta E_{jk}=E_j-E_k$.
Writing the survival probability and the transition probability separately, we obtain 
\begin{eqnarray}
P(\nu_{\alpha L}\to\nu_{\alpha L})&=&
1-\sum_{j<k}4|U_{\alpha j}U_{\alpha k}|^2 \sin^2 \left(\frac{\Delta E_{jk}t}{2}\right), \label{nonrela-survive}\\
P(\nu_{\alpha L}\to\nu_{\beta L})&=&-4\sum_{j<k}{\rm Re}[U_{\alpha j}U_{\beta j}^*U_{\alpha k}^*U_{\beta k}]\sin^2 \left(\frac{\Delta E_{jk}t}{2}\right)
-2\sum_{j<k}{\rm Im}[U_{\alpha j}U_{\beta j}^*U_{\alpha k}^*U_{\beta k}]\sin \Delta E_{jk}t. \label{nonrela-transition}
\end{eqnarray}
\end{widetext}
These representations are valid regardless of whether neutrinos are 
the Dirac particles or the Majorana particles and can be extended to $n$ generations.
In the usual oscillations without chirality-flip, the oscillation probability 
does not depend on the Majorana CP phase and depends only on the Dirac CP phase.

%%%%%%%%%%%%%%%%%%%%%%%%%%%%%%%%%%%%%%%%%%%%%%%%%%%%%%%%%%%
\section{Oscillation Probabilities of Dirac Neutrinos from Relativistic Equation}

In this section, we derive the neutrino oscillation probabilities from the relativistic equation
in the case of three-generation neutrinos with only Dirac mass term.
At first, we calculate the oscillation probabilities of left-handed neutrinos to other neutrinos.
Next, we investigate the CP dependence of these probabilities and check the unitarity.
Second, we calculate the probabilities of also left-handed but wrong-helicity neutrinos.
After that, we derive the probabilities of right-handed neutrinos and anti-neutrinos.

\subsection{Oscillation Probabilities of Left-Handed Neutrinos}

In three generations, the lagrangian for the Dirac neutrinos is represented by the spinors with four components as
\begin{eqnarray}
L&=&\displaystyle{\sum_{\alpha}(i\overline{\psi_{\alpha L}}\gamma^\mu \partial_\mu \psi_{\alpha L}
+i\overline{\psi_{\alpha R}}\gamma^\mu \partial_\mu \psi_{\alpha R})} \nonumber \\
&&-\displaystyle{\sum_{(\alpha,\beta)}\left(\overline{\psi_{\alpha L}}m_{\beta\alpha}^*\psi_{\beta R}
+\overline{\psi_{\alpha R}}m_{\alpha\beta}\psi_{\beta L}\right)},
\end{eqnarray}
where $(\alpha,\beta)$ means the sum over all combinations of $e$, $\mu$ and $\tau$.
The Eular-Lagrange equation for $\overline{\psi_{\alpha L}}$,
\begin{eqnarray}
\frac{\partial L}{\partial \overline{\psi_{\alpha L}}}
-\partial_\mu \left(\frac{\partial L}{\partial(\partial_\mu \overline{\psi_{\alpha L}})}\right)=0,
\end{eqnarray}
leads to the equation,
\begin{eqnarray}
i\gamma^\mu \partial_\mu \psi_{\alpha L}-\sum_{\beta}m_{\beta\alpha}^* \psi_{\beta R}=0.
\end{eqnarray}
Multiplying $\gamma^0$ from the left, the equiation becomes
\begin{eqnarray}
&&i\partial_0 \psi_{\alpha L}+i\gamma^0\gamma^i\partial_i \psi_{\alpha L}
-\sum_{\beta}m_{\beta\alpha}^* \gamma^0\psi_{\beta R}=0.
\end{eqnarray}
This equation is represented by two-component spinors $\xi$ and $\eta$ as
\begin{eqnarray}
&&\hspace{-0.5cm}i\partial_0 \left(\begin{array}{c}0 \\ \eta_{\alpha}\end{array}\right)+i\left(\begin{array}{cc}0 & 1 \\ 1 & 0\end{array}\right)
\left(\begin{array}{cc}0 & -\sigma_i \\ \sigma_i & 0\end{array}\right)\partial_i
\left(\begin{array}{c}0 \\ \eta_{\alpha}\end{array}\right) \nonumber \\
&&\hspace{1.5cm}-\sum_{\beta}m_{\beta\alpha}^* \left(\begin{array}{cc}0 & 1 \\ 1 & 0\end{array}\right)\left(\begin{array}{c}\xi_{\beta} \\ 0\end{array}\right)=0, \\
&&\hspace{-0.5cm}i\partial_0 \left(\begin{array}{c}0 \\ \eta_{\alpha}\end{array}\right)-i
\left(\begin{array}{c}0 \\ \sigma_i\partial_i\eta_{\alpha}\end{array}\right)
-\sum_{\beta}m_{\beta\alpha}^* \left(\begin{array}{c}0 \\ \xi_{\beta}\end{array}\right)=0.
\end{eqnarray}
Taking the lower two components, we obtain the equation,
\begin{eqnarray}
&&i\partial_0 \eta_{\alpha}-i\sigma_i\partial_i \eta_{\alpha}
-\sum_{\beta}m_{\beta\alpha}^* \xi_{\beta}=0. \label{eq-of-eta}
\end{eqnarray}
In the same way, the Eular-Lagrange equation
\begin{eqnarray}
\frac{\partial L}{\partial \overline{\psi_{\alpha R}}}
-\partial_\mu \left(\frac{\partial L}{\partial(\partial_\mu \overline{\psi_{\alpha R}})}\right)=0,
\end{eqnarray}
leads to
\begin{eqnarray}
i\gamma^\mu \partial_\mu \psi_{\alpha R}-\sum_{\beta}m_{\alpha\beta} \psi_{\beta L}=0.
\end{eqnarray}
Multiplying $\gamma^0$ from the left, the equation becomes
\begin{eqnarray}
&&i\partial_0 \psi_{\alpha R}+i\gamma^0\gamma^i\partial_i \psi_{\alpha R}
-\sum_{\beta}m_{\alpha\beta} \gamma^0\psi_{\beta L}=0.
\end{eqnarray}
This equation is also represented by two-component spinors, $\xi$ and $\eta$ as
\begin{eqnarray}
&&\hspace{-0.5cm}i\partial_0 \left(\begin{array}{c}\xi_{\alpha} \\ 0\end{array}\right)
+i\left(\begin{array}{cc}0 & 1 \\ 1 & 0\end{array}\right)
\left(\begin{array}{cc}0 & -\sigma_i \\ \sigma_i & 0\end{array}\right)\partial_i
\left(\begin{array}{c}\xi_{\alpha} \\ 0\end{array}\right) \nonumber \\
&&\hspace{1.5cm}-\sum_{\beta}m_{\alpha\beta} \left(\begin{array}{cc}0 & 1 \\ 1 & 0\end{array}\right)\left(\begin{array}{c}0 \\ \eta_{\beta} \end{array}\right)=0, \\
&&\hspace{-0.5cm}i\partial_0 \left(\begin{array}{c}\xi_{\alpha} \\ 0\end{array}\right)+i
\left(\begin{array}{c}\sigma_i\partial_i\xi_{\alpha} \\ 0\end{array}\right)
-\sum_{\beta}m_{\alpha\beta} \left(\begin{array}{c}\eta_{\beta} \\ 0\end{array}\right)=0.
\end{eqnarray}
Taking the upper two components, we obtain the equation,
\begin{eqnarray}
&&i\partial_0 \xi_{\alpha}+i\sigma_i\partial_i \xi_{\alpha}
-\sum_{\beta}m_{\alpha\beta} \eta_{\beta}=0. \label{eq-of-xi}
\end{eqnarray}
\begin{widetext}
Here, we assume the equal momentum for different flavors and factor out the dependence of the
distance as
\begin{eqnarray}
\eta_{\alpha}(x,t)&=&e^{i\vec{p}\cdot \vec{x}}\eta_{\alpha}(t)
=e^{i\vec{p}\cdot \vec{x}}\left(\begin{array}{c}\nu_{\alpha L}^{\prime} \\ \nu_{\alpha L}\end{array}\right),
\end{eqnarray}
\begin{eqnarray}
\xi_{\alpha}(x,t)&=&e^{i\vec{p}\cdot \vec{x}}\xi_{\alpha}(t)
=e^{i\vec{p}\cdot \vec{x}}\left(\begin{array}{c}\nu_{\alpha R}^{\prime} \\ \nu_{\alpha R}\end{array}\right). \label{equal-p}
\end{eqnarray}
Furthermore, if we choose $\vec{p}=(0,0,p)$, the equations (\ref{eq-of-eta}) and (\ref{eq-of-xi}) are rewritten as
\begin{eqnarray}
&&\hspace{-1cm}i\partial_0 \left(\begin{array}{c}\nu_{\alpha L}^{\prime} \\ \nu_{\alpha L}\end{array}\right)
+p\left(\begin{array}{c}\nu_{\alpha L}^{\prime} \\ -\nu_{\alpha L}\end{array}\right)
-\sum_{\beta}m_{\beta\alpha}^* \left(\begin{array}{c}\nu_{\beta R}^{\prime} \\ \nu_{\beta R}\end{array}\right)=0, \label{eq-nuL}
\end{eqnarray}
\begin{eqnarray}
&&\hspace{-1cm}i\partial_0 \left(\begin{array}{c}\nu_{\alpha R}^{\prime} \\ \nu_{\alpha R}\end{array}\right)
-p\left(\begin{array}{c}\nu_{\alpha R}^{\prime} \\ -\nu_{\alpha R}\end{array}\right)
-\sum_{\beta}m_{\alpha\beta} \left(\begin{array}{c}\nu_{\beta L}^{\prime} \\ \nu_{\beta L}\end{array}\right)=0. \label{eq-nuR}
\end{eqnarray}
If we write the equations (\ref{eq-nuL}) and (\ref{eq-nuR}) for three flavors together in a matrix form,
the time evolution of the chirality-flavor eigenstates is
represented by
%\begin{widetext}
\begin{eqnarray}
i\frac{d}{dt}\left(\begin{array}{c}
\nu_{eR}^{\prime} \\ \nu_{\mu R}^{\prime} \\ \nu_{\tau R}^{\prime} \\ \nu_{eL}^{\prime} \\ \nu_{\mu L}^{\prime} \\ \nu_{\tau L}^{\prime} \\
\nu_{eR} \\ \nu_{\mu R} \\ \nu_{\tau R} \\ \nu_{e L} \\ \nu_{\mu L} \\ \nu_{\tau L}
\end{array}\right)
=\left(\begin{array}{cccccc|cccccc}
p & 0 & 0 & m_{ee} & m_{e\mu} & m_{e\tau} & 0 & 0 & 0 & 0 & 0 & 0 \\
0 & p & 0 & m_{\mu e} & m_{\mu\mu} & m_{\mu\tau} & 0 & 0 & 0 & 0 & 0 & 0 \\
0 & 0 & p & m_{\tau e} & m_{\tau\mu} & m_{\tau\tau} & 0 & 0 & 0 & 0 & 0 & 0 \\
m_{ee}^* & m_{\mu e}^* & m_{\tau e}^* & -p & 0 & 0 & 0 & 0 & 0 & 0 & 0 & 0 \\
m_{e\mu}^* & m_{\mu\mu}^* & m_{\tau\mu}^* & 0 & -p & 0 & 0 & 0 & 0 & 0 & 0 & 0 \\
m_{e\tau}^* & m_{\mu\tau}^* & m_{\tau\tau}^* & 0 & 0 & -p & 0 & 0 & 0 & 0 & 0 & 0 \\
\hline
0 & 0 & 0 & 0 & 0 & 0 & -p & 0 & 0 & m_{ee} & m_{e\mu} & m_{e\tau} \\
0 & 0 & 0 & 0 & 0 & 0 & 0 & -p & 0 & m_{\mu e} & m_{\mu\mu} & m_{\mu\tau} \\
0 & 0 & 0 & 0 & 0 & 0 & 0 & 0 & -p & m_{\tau e} & m_{\tau\mu} & m_{\tau\tau} \\
0 & 0 & 0 & 0 & 0 & 0 & m_{ee}^* & m_{\mu e}^* & m_{\tau e}^* & p & 0 & 0 \\
0 & 0 & 0 & 0 & 0 & 0 & m_{e\mu}^* & m_{\mu\mu}^* & m_{\tau\mu}^* & 0 & p & 0 \\
0 & 0 & 0 & 0 & 0 & 0 & m_{e\tau}^* & m_{\mu\tau}^* & m_{\tau\tau}^* & 0 & 0 & p
\end{array}\right)
\left(\begin{array}{c}
\nu_{eR}^{\prime} \\ \nu_{\mu R}^{\prime} \\ \nu_{\tau R}^{\prime} \\ \nu_{eL}^{\prime} \\ \nu_{\mu L}^{\prime} \\ \nu_{\tau L}^{\prime} \\
\nu_{e R} \\ \nu_{\mu R} \\ \nu_{\tau R} \\ \nu_{e L} \\ \nu_{\mu L} \\ \nu_{\tau L}
\end{array}\right), \label{12-12-nu-matrix}
\end{eqnarray}
as same as the two generations.
As the lower-right part is separated from the upper-left part completely in the case of Dirac neutrinos,
they cannot mix each other even if the time has passed.
Below, we consider the lower-right part,
\begin{eqnarray}
i\frac{d}{dt}\left(\begin{array}{c}
\nu_{eR} \\ \nu_{\mu R} \\ \nu_{\tau R} \\ \nu_{e L} \\ \nu_{\mu L} \\ \nu_{\tau L}
\end{array}\right)
=\left(\begin{array}{ccc|ccc}
-p & 0 & 0 & m_{ee} & m_{e\mu} & m_{e\tau} \\
0 & -p & 0 & m_{\mu e} & m_{\mu\mu} & m_{\mu\tau} \\
0 & 0 & -p & m_{\tau e} & m_{\tau\mu} & m_{\tau\tau} \\
\hline
m_{ee}^* & m_{\mu e}^* & m_{\tau e}^* & p & 0 & 0 \\
m_{e\mu}^* & m_{\mu\mu}^* & m_{\tau\mu}^* & 0 & p & 0 \\
m_{e\tau}^* & m_{\mu\tau}^* & m_{\tau\tau}^* & 0 & 0 & p
\end{array}\right)
\left(\begin{array}{c}
\nu_{eR} \\ \nu_{\mu R} \\ \nu_{\tau R} \\ \nu_{e L} \\ \nu_{\mu L} \\ \nu_{\tau L}
\end{array}\right). \label{Dirac-Hamiltonian}
\end{eqnarray}
The chirality-flavor eigenstates are represented as the linear combination of the chirality-mass eigenstates,
\begin{eqnarray}
\left(\begin{array}{c}
\nu_{eR} \\ \nu_{\mu R} \\ \nu_{\tau R} \\ \nu_{e L} \\ \nu_{\mu L} \\ \nu_{\tau L}
\end{array}\right)
=\underbrace{\left(\begin{array}{ccc|ccc}
V_{e1} & V_{e2} & V_{e3} & 0 & 0 & 0 \\
V_{\mu 1} & V_{\mu 2} & V_{\mu 3} & 0 & 0 & 0 \\
V_{\tau 1} & V_{\tau 2} & V_{\tau 3} & 0 & 0 & 0 \\
\hline
0 & 0 & 0 & U_{e1} & U_{e2} & U_{e3} \\
0 & 0 & 0 & U_{\mu 1} & U_{\mu 2} & U_{\mu 3} \\
0 & 0 & 0 & U_{\tau 1} & U_{\tau 2} & U_{\tau 3}
\end{array}\right)}_{\cal U}\left(\begin{array}{c}
\nu_{1R} \\ \nu_{2R} \\ \nu_{3R} \\ \nu_{1L} \\ \nu_{2L} \\ \nu_{3L}
\end{array}\right). \label{cf-to-cm}
\end{eqnarray}
The mass submatrix in the Hamiltonian is diagonalized by the mixing matrix ${\cal U}$ defined above as
\begin{eqnarray}
{\cal U}^{\dagger}
\left(\begin{array}{ccc|ccc}
-p & 0 & 0 & m_{ee} & m_{e\mu} & m_{e\tau} \\
0 & -p & 0 & m_{\mu e} & m_{\mu\mu} & m_{\mu\tau} \\
0 & 0 & -p & m_{\tau e} & m_{\tau\mu} & m_{\tau\tau} \\
\hline
m_{ee}^* & m_{\mu e}^* & m_{\tau e}^* & p & 0 & 0 \\
m_{e\mu}^* & m_{\mu\mu}^* & m_{\tau\mu}^* & 0 & p & 0 \\
m_{e\tau}^* & m_{\mu\tau}^* & m_{\tau\tau}^* & 0 & 0 & p
\end{array}\right)
{\cal U}
=\left(\begin{array}{ccc|ccc}
-p & 0 & 0 & m_1 & 0 & 0 \\
0 & -p & 0 & 0 & m_2 & 0 \\
0 & 0 & -p & 0 & 0 & m_3 \\
\hline
m_1 & 0 & 0 & p & 0 & 0 \\
0 & m_2 & 0 & 0 & p & 0 \\
0 & 0 & m_3 & 0 & 0 & p \\
\end{array}\right).
\end{eqnarray}
The time evolution of the chirality-mass eigenstates is given by
\begin{eqnarray}
i\frac{d}{dt}\left(\begin{array}{c}
\nu_{1R} \\ \nu_{2R} \\ \nu_{3R} \\ \nu_{1L} \\ \nu_{2L} \\ \nu_{3L}
\end{array}\right)
=\left(\begin{array}{ccc|ccc}
-p & 0 & 0 & m_1 & 0 & 0 \\
0 & -p & 0 & 0 & m_2 & 0 \\
0 & 0 & -p & 0 & 0 & m_3 \\
\hline
m_1 & 0 & 0 & p & 0 & 0 \\
0 & m_2 & 0 & 0 & p & 0 \\
0 & 0 & m_3 & 0 & 0 & p \\
\end{array}\right)
\left(\begin{array}{c}
\nu_{1R} \\ \nu_{2R} \\ \nu_{3R} \\ \nu_{1L} \\ \nu_{2L} \\ \nu_{3L}
\end{array}\right). \label{eq-CMstate}
\end{eqnarray}
In order to diagonalize the Hamiltonian in eq.(\ref{eq-CMstate}) completely, let us rewrite this into the equation for the energy-helicity
eigenstates. Exchanging some rows and some columns of eq.(\ref{eq-CMstate}) and grouping by each generation,
it can be rewritten as
\begin{eqnarray}
i\frac{d}{dt}\left(\begin{array}{c}
\nu_{1R} \\ \nu_{1L} \\ \nu_{2R} \\ \nu_{2L} \\ \nu_{3R} \\ \nu_{3L}
\end{array}\right)
=\left(\begin{array}{cc|cc|cc}
-p & m_1 & 0 & 0 & 0 & 0 \\
m_1 & p & 0 & 0 & 0 & 0 \\\hline
0 & 0 & -p & m_2 & 0 & 0 \\
0 & 0 & m_2 & p & 0 & 0 \\\hline
0 & 0 & 0 & 0 & -p & m_3 \\
0 & 0 & 0 & 0 & m_3 & p
\end{array}\right)\left(\begin{array}{c}
\nu_{1R} \\ \nu_{1L} \\ \nu_{2R} \\ \nu_{2L} \\ \nu_{3R} \\ \nu_{3L}
\end{array}\right).
\end{eqnarray}
The chirality-mass eigenstates are represented as the linear combination of the energy-helicity eigenstates,
\begin{eqnarray}
\left(\begin{array}{c}
\nu_{1R} \\ \nu_{1L} \\ \nu_{2R} \\ \nu_{2L} \\ \nu_{3R} \\ \nu_{3L}
\end{array}\right)
=\underbrace{\left(\begin{array}{cc|cc|cc}
\sqrt{\frac{E_1+p}{2E_1}} & \sqrt{\frac{E_1-p}{2E_1}} & 0 & 0 & 0 & 0 \\
-\sqrt{\frac{E_1-p}{2E_1}} & \sqrt{\frac{E_1+p}{2E_1}} & 0 & 0 & 0 & 0 \\
\hline
0 & 0 & \sqrt{\frac{E_2+p}{2E_2}} & \sqrt{\frac{E_2-p}{2E_2}} & 0 & 0 \\
0 & 0 & -\sqrt{\frac{E_2-p}{2E_2}} & \sqrt{\frac{E_2+p}{2E_2}} & 0 & 0 \\
\hline
0 & 0 & 0 & 0 & \sqrt{\frac{E_3+p}{2E_3}} & \sqrt{\frac{E_3-p}{2E_3}} \\
0 & 0 & 0 & 0 & -\sqrt{\frac{E_3-p}{2E_3}} & \sqrt{\frac{E_3+p}{2E_3}}
\end{array}\right)}_{W}\left(\begin{array}{c}
\nu_{1}^- \\ \nu_{1}^+ \\ \nu_{2}^- \\ \nu_{2}^+ \\ \nu_{3}^- \\ \nu_{3}^+
\end{array}\right). \label{chi-ene}
\end{eqnarray}
Then, the time evolution of the energy-helicity eigenstates is given by
\begin{eqnarray}
i\frac{d}{dt}\left(\begin{array}{c}
\nu_{1}^- \\ \nu_{1}^+ \\ \nu_{2}^- \\ \nu_{2}^+ \\ \nu_{3}^- \\ \nu_{3}^+
\end{array}\right)
=\left(\begin{array}{cccccc}
-E_1 & 0 & 0 & 0 & 0 & 0 \\
0 & E_1 & 0 & 0 & 0 & 0 \\
0 & 0 & -E_2 & 0 & 0 & 0 \\
0 & 0 & 0 & E_2 & 0 & 0 \\
0 & 0 & 0 & 0 & -E_3 & 0 \\
0 & 0 & 0 & 0 & 0 & E_3
\end{array}\right)\left(\begin{array}{c}
\nu_{1}^- \\ \nu_{1}^+ \\ \nu_{2}^- \\ \nu_{2}^+ \\ \nu_{3}^- \\ \nu_{3}^+
\end{array}\right), \label{time-evo1}
\end{eqnarray}
where
\begin{eqnarray}
E_j=\sqrt{p^2+m_j^2} \qquad (j=1,2,3).
\end{eqnarray}
Connecting eqs.(\ref{cf-to-cm}) and (\ref{chi-ene}), the chirality-flavor eigenstates are represented by the energy-helicity eigenstates,
\begin{eqnarray}
\left(\begin{array}{c}
\nu_{eR} \\ \nu_{\mu R} \\ \nu_{\tau R} \\ \nu_{e L} \\ \nu_{\mu L} \\ \nu_{\tau L}
\end{array}\right)
&=&
{\cal U}
\left(\begin{array}{cccccc}
1 & 0 & 0 & 0 & 0 & 0 \\
0 & 0 & 1 & 0 & 0 & 0 \\
0 & 0 & 0 & 0 & 1 & 0 \\
0 & 1 & 0 & 0 & 0 & 0 \\
0 & 0 & 0 & 1 & 0 & 0 \\
0 & 0 & 0 & 0 & 0 & 1
\end{array}\right)
W\left(\begin{array}{c}
\nu_{1}^- \\ \nu_{1}^+ \\ \nu_{2}^- \\ \nu_{2}^+ \\ \nu_{3}^- \\ \nu_{3}^+
\end{array}\right) \nonumber \\
&=&\left(\begin{array}{cc|cc|cc}
\sqrt{\frac{E_1+p}{2E_1}}V_{e 1} & \sqrt{\frac{E_1-p}{2E_1}}V_{e 1}
& \sqrt{\frac{E_2+p}{2E_2}}V_{e 2} & \sqrt{\frac{E_2-p}{2E_2}}V_{e 2}
& \sqrt{\frac{E_3+p}{2E_3}}V_{e 3} & \sqrt{\frac{E_3-p}{2E_3}}V_{e 3}\\
\sqrt{\frac{E_1+p}{2E_1}}V_{\mu 1} & \sqrt{\frac{E_1-p}{2E_1}}V_{\mu 1}
& \sqrt{\frac{E_2+p}{2E_2}}V_{\mu 2} & \sqrt{\frac{E_2-p}{2E_2}}V_{\mu 2}
& \sqrt{\frac{E_3+p}{2E_3}}V_{\mu 3} & \sqrt{\frac{E_3-p}{2E_3}}V_{\mu 3} \\
\sqrt{\frac{E_1+p}{2E_1}}V_{\tau 1} & \sqrt{\frac{E_1-p}{2E_1}}V_{\tau 1}
& \sqrt{\frac{E_2+p}{2E_2}}V_{\tau 2} & \sqrt{\frac{E_2-p}{2E_2}}V_{\tau 2}
& \sqrt{\frac{E_3+p}{2E_3}}V_{\tau 3} & \sqrt{\frac{E_3-p}{2E_3}}V_{\tau 3} \\
\hline
-\sqrt{\frac{E_1-p}{2E_1}}U_{e 1} & \sqrt{\frac{E_1+p}{2E_1}}U_{e 1}
& -\sqrt{\frac{E_2-p}{2E_2}}U_{e 2} & \sqrt{\frac{E_2+p}{2E_2}}U_{e 2}
& -\sqrt{\frac{E_3-p}{2E_3}}U_{e 3} & \sqrt{\frac{E_3+p}{2E_3}}U_{e 3} \\
-\sqrt{\frac{E_1-p}{2E_1}}U_{\mu 1} & \sqrt{\frac{E_1+p}{2E_1}}U_{\mu 1}
& -\sqrt{\frac{E_2-p}{2E_2}}U_{\mu 2} & \sqrt{\frac{E_2+p}{2E_2}}U_{\mu 2}
& -\sqrt{\frac{E_3-p}{2E_3}}U_{\mu 3} & \sqrt{\frac{E_3+p}{2E_3}}U_{\mu 3} \\
-\sqrt{\frac{E_1-p}{2E_1}}U_{\tau 1} & \sqrt{\frac{E_1+p}{2E_1}}U_{\tau 1}
& -\sqrt{\frac{E_2-p}{2E_2}}U_{\tau 2} & \sqrt{\frac{E_2+p}{2E_2}}U_{\tau 2}
& -\sqrt{\frac{E_3-p}{2E_3}}U_{\tau 3} & \sqrt{\frac{E_3+p}{2E_3}}U_{\tau 3}
\end{array}\right)
\left(\begin{array}{c}
\nu_{1}^- \\ \nu_{1}^+ \\ \nu_{2}^- \\ \nu_{2}^+ \\ \nu_{3}^- \\ \nu_{3}^+
\end{array}\right).
\label{flavor-energy}
\end{eqnarray}
%\end{widetext}
As same as the non-relativistic case, rewriting the relation about the fields to one particle states,
we obtain
\begin{eqnarray}
|\nu_{\alpha L}(t)\rangle=\sum_{j=1}^3\left(-\sqrt{\frac{E_j-p}{2E_j}}U_{\alpha j}^*e^{iE_jt}|\nu_j^-\rangle
+\sqrt{\frac{E_j+p}{2E_j}}U_{\alpha j}^*e^{-iE_jt}|\nu_j^+\rangle\right),
\end{eqnarray}
and their conjugate states,
%\begin{widetext}
\begin{eqnarray}
\langle \nu_{\beta L}| &=&\sum_{j=1}^3\left(-\sqrt{\frac{E_j-p}{2E_j}}U_{\beta j}\langle \nu_j^-|
+\sqrt{\frac{E_j+p}{2E_j}}U_{\beta j}\langle \nu_j^+|\right) \\
\langle \nu_{\beta R}| &=&\sum_{j=1}^3\left(\sqrt{\frac{E_j+p}{2E_j}}V_{\beta j}\langle \nu_j^-|
+\sqrt{\frac{E_j-p}{2E_j}}V_{\beta j}\langle \nu_j^+|\right).
\end{eqnarray}
Note that a one particle state of $\nu_{\alpha L}$ includes not only the positive energy parts
but also the negative energy parts.
This is the difference between the relativistic and non-relativistic case.
In the same way as the non-relativistic case, we calculate the amplitudes,
%\begin{widetext}
\begin{eqnarray}
A(\nu_{\alpha L}\to\nu_{\beta L})&=&\langle \nu_{\beta L}|\nu_{\alpha L}(t)\rangle
=\sum_j U_{\alpha j}^*U_{\beta j}\left(\frac{E_j-p}{2E_j}e^{iE_jt}+\frac{E_j+p}{2E_j}e^{-iE_jt}\right) \nonumber \\
&&\hspace{-1cm}=\sum_j U_{\alpha j}^*U_{\beta j}\left(\frac{e^{iE_jt}+e^{-iE_jt}}{2}-\frac{p}{E_j}\frac{e^{iE_jt}-e^{-iE_jt}}{2}\right)
=\sum_j U_{\alpha j}^*U_{\beta j}\left\{\cos (E_jt)-i\frac{p}{E_j}\sin (E_jt)\right\}, \\
A(\nu_{\alpha L}\to\nu_{\beta R})&=&\langle \nu_{\beta R}|\nu_{\alpha L}(t)\rangle
=\sum_j -U_{\alpha j}^*V_{\beta j}\frac{m_j}{2E_j}(e^{iE_jt}-e^{-iE_jt})
=\sum_j -iU_{\alpha j}^*V_{\beta j}\frac{m_j}{E_j}\sin (E_jt).
\end{eqnarray}
Furthermore, we derive the oscillation probabilities by squaring the absolute value of the amplitudes,
\begin{eqnarray}
P(\nu_{\alpha L}\to\nu_{\beta L})&=&
\sum_j |U_{\alpha j}^*U_{\beta j}|^2\left\{1-\frac{E_j^2-p^2}{E_j^2}\sin^2 (E_jt)\right\}\nonumber \\
&&\hspace{-1.3cm}+2\sum_{j<k}{\rm Re}\left[U_{\alpha j}^*U_{\beta j}U_{\alpha k}U_{\beta k}^*\right]
\left\{\cos (\Delta E_{jk}t)-\frac{E_jE_k-p^2}{E_jE_k}\sin (E_jt)\sin (E_kt)\right\} \nonumber \\
&&\hspace{-1.3cm}+2\sum_{j<k}{\rm Im}\left[U_{\alpha j}^*U_{\beta j}U_{\alpha k}U_{\beta k}^*\right]
\left\{\sin (\Delta E_{jk}t)+\frac{E_k-p}{E_k}\cos (E_jt)\sin (E_kt)-\frac{E_j-p}{E_j}\cos (E_kt)\sin (E_jt)\right\} \nonumber \\
&&\hspace{-1.3cm}=
\delta_{\alpha\beta}-\sum_j |U_{\alpha j}U_{\beta j}|^2\left\{\frac{m_j^2}{E_j^2}\sin^2 (E_jt)\right\}\nonumber \\
&&\hspace{-1.3cm}-2\sum_{j<k}{\rm Re}\left[U_{\alpha j}U_{\beta j}^*U_{\alpha k}^*U_{\beta k}\right]
\left\{2\sin^2 \left(\frac{\Delta E_{jk}t}{2}\right)+\frac{E_jE_k-p^2}{E_jE_k}\sin (E_jt)\sin (E_kt)\right\} \nonumber \\
&&\hspace{-1.3cm}-2\sum_{j<k}{\rm Im}\left[U_{\alpha j}U_{\beta j}^*U_{\alpha k}^*U_{\beta k}\right]
\left\{\sin (\Delta E_{jk}t)+\frac{E_k-p}{E_k}\cos (E_jt)\sin (E_kt)-\frac{E_j-p}{E_j}\cos (E_kt)\sin (E_jt)\right\}, \\
P(\nu_{\alpha L}\to\nu_{\beta R})&=&
\sum_j |U_{\alpha j}V_{\beta j}|^2\frac{m_j^2}{E_j^2}\sin^2 (E_jt)
+2\sum_{j<k}{\rm Re}[U_{\alpha j}U_{\alpha k}^*V_{\beta j}^*V_{\beta k}]\frac{m_jm_k}{E_jE_k}\sin (E_jt)\sin (E_kt).
\end{eqnarray}
If we describe the survival probabilities (the case for $\alpha=\beta$) and the transition probabilities
(the case for $\alpha\neq \beta$) separately, we obtain
\begin{eqnarray}
P(\nu_{\alpha L}\to\nu_{\alpha L})&=&
1-\sum_j |U_{\alpha j}|^4\left\{\frac{m_j^2}{E_j^2}\sin^2 (E_jt)\right\}\label{rela-survive}\\
&&\hspace{-1cm}-2\sum_{j<k}\left|U_{\alpha j}U_{\alpha k}\right|^2
\left\{2\sin^2 \left(\frac{\Delta E_{jk}t}{2}\right)+\frac{E_jE_k-p^2}{E_jE_k}\sin (E_jt)\sin (E_kt)\right\},
\label{rela-survive2}\\
P(\nu_{\alpha L}\to\nu_{\beta L})&=&
-\sum_j |U_{\alpha j}U_{\beta j}|^2\left\{\frac{m_j^2}{E_j^2}\sin^2 (E_jt)\right\}\label{rela-transition1}\\
&&\hspace{-1cm}-2\sum_{j<k}{\rm Re}\left[U_{\alpha j}U_{\beta j}^*U_{\alpha k}^*U_{\beta k}\right]
\left\{2\sin^2 \left(\frac{\Delta E_{jk}t}{2}\right)+\frac{E_jE_k-p^2}{E_jE_k}\sin (E_jt)\sin (E_kt)\right\} \label{rela-transition2}\\
&&\hspace{-1cm}-2\sum_{j<k}{\rm Im}\left[U_{\alpha j}U_{\beta j}^*U_{\alpha k}^*U_{\beta k}\right]
\left\{\sin (\Delta E_{jk}t)+\frac{E_k-p}{E_k}\cos (E_jt)\sin (E_kt)-\frac{E_j-p}{E_j}\cos (E_kt)\sin (E_jt)\right\},
\label{rela-transition}\\
P(\nu_{\alpha L}\to\nu_{\alpha R})&=&
\sum_j |U_{\alpha j}V_{\alpha j}|^2\frac{m_j^2}{E_j^2}\sin^2 (E_jt)
+2\sum_{j<k}{\rm Re}[U_{\alpha j}U_{\alpha k}^*V_{\alpha j}^*V_{\alpha k}]\frac{m_jm_k}{E_jE_k}\sin (E_jt)\sin (E_kt),
\label{chirality-change1}\\
P(\nu_{\alpha L}\to\nu_{\beta R})&=&
\sum_j |U_{\alpha j}V_{\beta j}|^2\frac{m_j^2}{E_j^2}\sin^2 (E_jt)
+2\sum_{j<k}{\rm Re}[U_{\alpha j}U_{\alpha k}^*V_{\beta j}^*V_{\beta k}]\frac{m_jm_k}{E_jE_k}\sin (E_jt)\sin (E_kt).
\label{chirality-change2}
\end{eqnarray}
\end{widetext}
In three or more generations, the terms proportional to the imaginary part of the product for four matrix elements represented in (\ref{rela-transition}) is added.
The results, (\ref{rela-survive})-(\ref{rela-transition}), derived by the relativistic method
should be compared by the results, (\ref{nonrela-survive}) and (\ref{nonrela-transition}), derived by
the non-relativistic method.
Comparing these equations, one can see that the first terms of (\ref{rela-survive}), (\ref{rela-survive2}), (\ref{rela-transition2})
and (\ref{rela-transition}) are equal to the results by the non-relativistic method and
the remaining terms is new terms appeared as the correction.
It is also noted that the representation of the above probabilities is parameter independent of the unitary matrix.

%%%%%%%%%%%%%%%%%%%%%%%%%%%%%%%%%
\subsection{CP dependence of Oscillation probability}

Next, let us consider the CP dependence of the oscillation probabilities.
We would like to show how the new CP phases appear with chirality-flip in the oscillation probabilities.
The $3\times 3$ unitary matrix is in general represented by nine parameters as
\begin{eqnarray}
\hspace{-0.5cm}
U&=&\left(\begin{array}{ccc}
e^{i\rho_{eL}} & 0\\
0 & e^{i\rho_{\mu L}} & 0 \\
0 & 0 & e^{i\rho_{\tau L}}
\end{array}\right)
\tilde{U}
\left(\begin{array}{ccc}
1 & 0 & 0 \\
0 & e^{i\phi_{2L}} & 0 \\
0 & 0 & e^{\phi_{3L}}
\end{array}\right),
\label{U}
\end{eqnarray}
\begin{eqnarray}
\hspace{-0.5cm}
V&=&\left(\begin{array}{ccc}
e^{i\rho_{eR}} & 0 & 0\\
0 & e^{i\rho_{\mu R}} & 0 \\
0 & 0 & e^{i\rho_{\tau R}}
\end{array}\right)
\tilde{V}
\left(\begin{array}{ccc}
1 & 0 & 0 \\
0 & e^{i\phi_{2R}} & 0 \\
0 & 0 & e^{i\phi_{3R}}
\end{array}\right), \label{V}
\end{eqnarray}
where $\tilde{U}$ and $\tilde{V}$ are the Maki-Nakagawa-Sakata (MNS) matrix
for left-handed neutrinos and right-handed neutrinos respectively.
Each MNS matrix is represented by three mixing angles and one CP phase in the case of three generations.
We can choose $\phi_{1L}=\phi_{1R}=0$ without loss of generality in the above equation.
Using (\ref{U}) and (\ref{V}), we can calculate the product of four matrix elements in the probabilities
(\ref{rela-survive})-(\ref{rela-transition}),
\begin{widetext}
\begin{eqnarray}
&&U_{\alpha j}U_{\beta j}^*U_{\alpha k}^*U_{\beta k}
=e^{i\rho_{\alpha L}}\tilde{U}_{\alpha j}e^{i\phi_{jL}}
e^{-i\rho_{\beta L}}\tilde{U}_{\beta j}^*e^{-i\phi_{jL}}
e^{-i\rho_{\alpha L}}\tilde{U}_{\alpha k}^*e^{-i\phi_{kL}}
e^{i\rho_{\beta L}}\tilde{U}_{\beta k}e^{i\phi_{kL}}
%\\&&\hspace{2.4cm}
=\tilde{U}_{\alpha j}\tilde{U}_{\beta j}^*
\tilde{U}_{\alpha k}^*\tilde{U}_{\beta k} \\
&&\hspace{0.1cm}V_{\alpha j}V_{\beta j}^*V_{\alpha k}^*V_{\beta k}
=e^{i\rho_{\alpha R}}\tilde{V}_{\alpha j}e^{i\phi_{jR}}
e^{-i\rho_{\beta R}}\tilde{V}_{\beta j}^*e^{-i\phi_{jR}}
e^{-i\rho_{\alpha R}}\tilde{V}_{\alpha k}^*e^{-i\phi_{kR}}
e^{i\rho_{\beta R}}\tilde{V}_{\beta k}e^{i\phi_{kR}}
%\\&&\hspace{2.4cm}
=\tilde{V}_{\alpha j}\tilde{V}_{\beta j}^*
\tilde{V}_{\alpha k}^*\tilde{V}_{\beta k}
\end{eqnarray}
One can see that the overall phases cancel out and only the Dirac CP phases included in $\tilde{U}$ and
$\tilde{V}$ remain in the oscillation probabilities without chirality-flip.
On the other hand, the product of four matrix elements in the probabilities
(\ref{chirality-change1}) and (\ref{chirality-change2}) becomes
\begin{eqnarray}
&&U_{\alpha j}U_{\alpha k}^*V_{\beta j}^*V_{\beta k}
=e^{i\rho_{\alpha L}}\tilde{U}_{\alpha j}e^{i\phi_{jL}}
e^{-i\rho_{\alpha L}}\tilde{U}_{\alpha k}^*e^{-i\phi_{kL}}
e^{-i\rho_{\beta R}}\tilde{V}_{\beta j}^*e^{-i\phi_{jR}}
e^{i\rho_{\beta R}}\tilde{V}_{\beta k}e^{i\phi_{kR}} \nonumber \\
&&\hspace{2.2cm}
=\tilde{U}_{\alpha j}\tilde{U}_{\alpha k}^*
\tilde{V}_{\beta j}^* \tilde{V}_{\beta k}
e^{i(\phi_{jL}-\phi_{kL}-\phi_{jR}+\phi_{kR})}.
\end{eqnarray}
\end{widetext}
The dependence of overall phases remains in the oscillation probabilities with chirality-flip.
In principle, these overall phases can be observed if we can distinguish
the flavor of right-handed neutrinos beyond the Standard Model and then these phases become new CP phases.
The probabilities, (\ref{chirality-change1}) and (\ref{chirality-change2}) depend on the new CP phases only through
${\rm Re}\left[U_{\alpha j}U_{\alpha k}^*V_{\beta j}^*V_{\beta k}\right]$.
Therefore, the effect of the new CP phases can be measured indirectly from the oscillation probabilities
and there is no direct CP violation related to the new CP phases, namely
the difference, $P(\nu_{\alpha L}\to\nu_{\beta R})-P(\nu_{\beta R}\to\nu_{\alpha L})$ vanishes.
However, the real part of the product of four mixing matrix elements is decomposed as
%\begin{widetext}
\begin{eqnarray}
&&\hspace{-0.8cm}
%\hspace{-1cm}
{\rm Re}\left[U_{\alpha j}U_{\alpha k}^*V_{\beta j}^*V_{\beta k}\right]\nonumber \\
&&\hspace{-0.8cm}={\rm Re}\left[\tilde{U}_{\alpha j}\tilde{U}_{\alpha k}^*
\tilde{V}_{\beta j}^*\tilde{V}_{\beta k}
e^{i(\phi_{jL}-\phi_{kL}-\phi_{jR}+\phi_{kR})}\right] \nonumber \\
&&\hspace{-0.8cm}
%\hspace{0.2cm}
={\rm Re}\left[\tilde{U}_{\alpha j}\tilde{U}_{\alpha k}^*
\tilde{V}_{\beta j}^*\tilde{V}_{\beta k}\right]
\cos(\phi_{jL}-\phi_{kL}-\phi_{jR}+\phi_{kR})\nonumber \\
&&\hspace{-0.8cm}-{\rm Im}\left[\tilde{U}_{\alpha j}\tilde{U}_{\alpha k}^*
\tilde{V}_{\beta j}^* \tilde{V}_{\beta k}\right]
\sin(\phi_{jL}-\phi_{kL}-\phi_{jR}+\phi_{kR}).
\end{eqnarray}
If the matrix elements of $\tilde{U}$ and $\tilde{V}$ are investigated by the oscillations of left-handed and right-handed 
neutrinos, we can obtain the information of the new CP phases through both cosine and sine terms.
The cosine term alone cannot determine the value of a new CP phase as one of 360 degrees,
but it can be determined by measuring both cosine and sine terms.

Let us count the number of independent parameters related to the new CP phases.
If we define $\Delta \phi_{jkL}=\phi_{jL}-\phi_{kL}$ and $\Delta \phi_{jkR}=\phi_{jR}-\phi_{kR}$,
the probabilities depend through the form of $\Delta \phi_{jkL}-\Delta \phi_{jkR}$,
where both $j$ and $k$ run from 1 to 3 in three generations.
As the relation $\Delta \phi_{13L}=\Delta \phi_{12L}-\Delta \phi_{23L}$ etc. holds for example,
the number of independent parameters including the new CP phases is two in three generations.
Extending the above discussion to the case of $n$ generations, the number of new phases becomes
\begin{eqnarray}
n-1,
\end{eqnarray}
and the number of the CP phases included in the MNS matrix is
\begin{eqnarray}
\frac{(n-1)(n-2)}{2}
\end{eqnarray}
as in the previous case.
Summing up these two kinds of phases, the total number of the CP phases is
\begin{eqnarray}
\frac{n(n-1)}{2}.
\end{eqnarray}

%%%%%%%%%%%%%%%%%%%%%%%%%%%%
\subsection{Unitary Check of Oscillation Probabilities}

Next, let us confirm the unitarity in the framework of three generations.
In the Standard Model, a right-handed neutrino can be chosen as the mass eigenstate
because $\nu_R$ does not interact through weak interactions.
In this case, the matrix $V$, which mixes right-handed neutrinos, becomes the unit matrix
and the mixing angles and the CP phase corresponding to $\nu_R$ do not appear.
Then, only the sum of the oscillation probabilities for $\nu_{\alpha L}\to \nu_R$,
\begin{widetext}
\begin{eqnarray}
P(\nu_{\alpha L}\to\nu_{R})&=&\sum_{\beta} P(\nu_{\alpha L}\to\nu_{\beta R})\nonumber \\
&=&\sum_{\beta} \sum_j |U_{\alpha j}V_{\beta j}|^2\frac{m_j^2}{E_j^2}\sin^2 (E_jt)
+2\sum_{\beta}\sum_{j<k}{\rm Re}[U_{\alpha j}U_{\alpha k}^*V_{\beta j}^*V_{\beta k}]\frac{m_jm_k}{E_jE_k}\sin (E_jt)\sin (E_kt)
\nonumber \\
&=&\sum_j |U_{\alpha j}|^2\frac{m_j^2}{E_j^2}\sin^2 (E_jt),\label{LtoR}
\end{eqnarray}
is observable, where we use the unitarity of $V$.
We can also calculate the sum of the probabilities for $\nu_{\alpha L} \to \nu_L$,
\begin{eqnarray}
P(\nu_{\alpha L}\to\nu_{L})&=&\sum_{\beta} P(\nu_{\alpha L}\to\nu_{\beta L})\nonumber \\
&&\hspace{-1.5cm}=\sum_{\beta}\delta_{\alpha\beta}-\sum_{\beta}\sum_j |U_{\alpha j}U_{\beta j}|^2\left\{\frac{m_j^2}{E_j^2}\sin^2 (E_jt)\right\}\nonumber \\
&&\hspace{-1.5cm}-2\sum_{\beta}\sum_{j<k}{\rm Re}\left[U_{\alpha j}U_{\beta j}^*U_{\alpha k}^*U_{\beta k}\right]
\left\{2\sin^2 \left(\frac{\Delta E_{jk}t}{2}\right)+\frac{E_jE_k-p^2}{E_jE_k}\sin (E_jt)\sin (E_kt)\right\} \nonumber \\
&&\hspace{-1.5cm}+2\sum_{\beta}\sum_{j<k}{\rm Im}\left[U_{\alpha j}U_{\beta j}^*U_{\alpha k}^*U_{\beta k}\right]
\left\{\sin (\Delta E_{jk}t)+\frac{E_k-p}{E_k}\cos (E_jt)\sin (E_kt)-\frac{E_j-p}{E_j}\cos (E_kt)\sin (E_jt)\right\} \nonumber \\
&&\hspace{-1.5cm}=1-\sum_j |U_{\alpha j}|^2\left\{\frac{m_j^2}{E_j^2}\sin^2 (E_jt)\right\},\label{LtoL}
\end{eqnarray}
and adding (\ref{LtoR}) and (\ref{LtoL}), we obtain
\begin{eqnarray}
P(\nu_{\alpha L}\to\nu_{R})+P(\nu_{\alpha L}\to\nu_{L})=1,
\end{eqnarray}
and therefore we have confirmed the unitarity by adding all probabilities with and without chirality-flip.

We can also confirm the unitarity for a right-handed neutrino by replacing $U\leftrightarrow V$,
for an anti-neutrino by replacing $U\rightarrow U^*$, $V\rightarrow V^*$
and for a Majorana neutrino by replacing $V \rightarrow U^*$ in equations, (\ref{LtoR})-(\ref{LtoL}).

%%%%%%%%%%%%%%%%%%%%%%%%%%%%%
\subsection{Oscillation Probabilities of Wrong-Helicity Neutrinos}

Next, let us consider the upper-left part of (\ref{12-12-nu-matrix}), namely
the oscillation probabilities for $\nu^{\prime}$.
The time evolution of the chirality-flavor eigenstates is given by
\begin{eqnarray}
i\frac{d}{dt}\left(\begin{array}{c}
\nu_{eR}^{\prime} \\ \nu_{\mu R}^{\prime} \\ \nu_{\tau R}^{\prime} \\
\nu_{e L}^{\prime} \\ \nu_{\mu L}^{\prime} \\ \nu_{\tau L}^{\prime}
\end{array}\right)
=\left(\begin{array}{ccc|ccc}
p & 0 & 0 & m_{ee} & m_{e\mu} & m_{e\tau} \\
0 & p & 0 & m_{\mu e} & m_{\mu\mu} & m_{\mu\tau} \\
0 & 0 & p & m_{\tau e} & m_{\tau\mu} & m_{\tau\tau} \\
\hline
m_{ee}^* & m_{\mu e}^* & m_{\tau e}^* & -p & 0 & 0 \\
m_{e\mu}^* & m_{\mu\mu}^* & m_{\tau\mu}^* & 0 & -p & 0 \\
m_{e\tau}^* & m_{\mu\tau}^* & m_{\tau\tau}^* & 0 & 0 & -p
\end{array}\right)
\left(\begin{array}{c}
\nu_{eR}^{\prime} \\ \nu_{\mu R}^{\prime} \\ \nu_{\tau R}^{\prime} \\
\nu_{e L}^{\prime} \\ \nu_{\mu L}^{\prime} \\ \nu_{\tau L}^{\prime}
\end{array}\right). \label{Dirac-Hamiltonian2}
\end{eqnarray}
Comparing (\ref{Dirac-Hamiltonian2}) to (\ref{Dirac-Hamiltonian}),
we can see that the sign of $p$ in the Hamiltonian is reversed from the case for $\nu$.
Therefore, the probabilities of $\nu^{\prime}$ are obtained by changing the sign of $p$
in eqs.(\ref{rela-survive})-(\ref{chirality-change2}).
As changing the sign of $p$ does not change the energy $E_j=\sqrt{p^2+m_j^2}$,
the probabilities except for (\ref{rela-transition}) do not change.
On the other hand, the eq.(\ref{rela-transition}) can be expressed as
\begin{eqnarray}
&&\hspace{-1cm}-2\sum_{j<k}{\rm Im}\left[U_{\alpha j}U_{\beta j}^*U_{\alpha k}^*U_{\beta k}\right]
\left\{\sin (\Delta E_{jk}t)+\frac{E_k-p}{E_k}\cos (E_jt)\sin (E_kt)-\frac{E_j-p}{E_j}\cos (E_kt)\sin (E_jt)\right\}, \nonumber \\
&&\hspace{-1cm}=2p\sum_{j<k}{\rm Im}\left[U_{\alpha j}U_{\beta j}^*U_{\alpha k}^*U_{\beta k}\right]
\left\{\frac{1}{E_k}\cos (E_jt)\sin (E_kt)-\frac{1}{E_j}\cos (E_kt)\sin (E_jt)\right\},
\end{eqnarray}
and is proportional to the momentum $p$.
Namely, the sign of (\ref{rela-transition}) changes according to reversing the sign of $p$.
As a result, the probabilities of $\nu^{\prime}$ are given by
\begin{eqnarray}
P(\nu_{\alpha L}^{\prime}\to\nu_{\alpha L}^{\prime})&=&
1-\sum_j |U_{\alpha j}|^4\left\{\frac{m_j^2}{E_j^2}\sin^2 (E_jt)\right\}\label{rela-survive-prime}\nonumber \\
&&\hspace{-1cm}-2\sum_{j<k}\left|U_{\alpha j}U_{\alpha k}\right|^2
\left\{2\sin^2 \left(\frac{\Delta E_{jk}t}{2}\right)+\frac{E_jE_k-p^2}{E_jE_k}\sin (E_jt)\sin (E_kt)\right\},
\\
P(\nu_{\alpha L}^{\prime}\to\nu_{\beta L}^{\prime})&=&
-\sum_j |U_{\alpha j}U_{\beta j}|^2\left\{\frac{m_j^2}{E_j^2}\sin^2 (E_jt)\right\}\nonumber \\
&&\hspace{-1cm}-2\sum_{j<k}{\rm Re}\left[U_{\alpha j}U_{\beta j}^*U_{\alpha k}^*U_{\beta k}\right]
\left\{2\sin^2 \left(\frac{\Delta E_{jk}t}{2}\right)+\frac{E_jE_k-p^2}{E_jE_k}\sin (E_jt)\sin (E_kt)\right\} \nonumber \\
&&\hspace{-1cm}+2\sum_{j<k}{\rm Im}\left[U_{\alpha j}U_{\beta j}^*U_{\alpha k}^*U_{\beta k}\right]
\left\{\sin (\Delta E_{jk}t)+\frac{E_k-p}{E_k}\cos (E_jt)\sin (E_kt)-\frac{E_j-p}{E_j}\cos (E_kt)\sin (E_jt)\right\},
\label{rela-transition-prime}\\
P(\nu_{\alpha L}^{\prime}\to\nu_{\alpha R}^{\prime})&=&
\sum_j |U_{\alpha j}V_{\alpha j}|^2\frac{m_j^2}{E_j^2}\sin^2 (E_jt)
+2\sum_{j<k}{\rm Re}[U_{\alpha j}U_{\alpha k}^*V_{\alpha j}^*V_{\alpha k}]\frac{m_jm_k}{E_jE_k}\sin (E_jt)\sin (E_kt),\\
P(\nu_{\alpha L}^{\prime}\to\nu_{\beta R}^{\prime})&=&
\sum_j |U_{\alpha j}V_{\beta j}|^2\frac{m_j^2}{E_j^2}\sin^2 (E_jt)
+2\sum_{j<k}{\rm Re}[U_{\alpha j}U_{\alpha k}^*V_{\beta j}^*V_{\beta k}]\frac{m_jm_k}{E_jE_k}\sin (E_jt)\sin (E_kt).
\label{naLp-nbRp}
\end{eqnarray}
%\end{widetext}
It is noted that in the oscillation probabilities for $\nu_{\alpha L}^{\prime}$ the sign of the term proportional to
${\rm Im}\left[U_{\alpha j}U_{\beta j}^*U_{\alpha k}^*U_{\beta k}\right]$ is reversed to those for $\nu_{\alpha L}$
although $\nu_{\alpha L}$ and $\nu_{\alpha L}^{\prime}$ have the same chirality.
This is because the Hamiltonians in (\ref{Dirac-Hamiltonian}) and (\ref{Dirac-Hamiltonian2}) have opposite sign
of momentum.

It is noted that the oscillation probabilities for $\nu$ and $\nu^{\prime}$ are
different in three or more generations.
To the best of our knowledge, this is the first paper to point this out.

%%%%%%%%%%%%%%%%%%%%%%%%%%
\subsection{Oscillation Probabilities of Right-Handed Neutrinos}

Next, we consider the oscillation of right-handed neutrinos.
If the flavor of the right-handed neutrinos can be distinguished beyond the Standard Model,
the oscillations of the right-handed neutrinos can be occured and we can calculate the oscillation probabilities.
For completeness, we describe these probabilities.
In the same way as the case for left-handed neutrinos, from (\ref{flavor-energy}), the amplitudes are given by
%\begin{widetext}
\begin{eqnarray}
A(\nu_{\alpha R}\to\nu_{\beta R})&=&
\sum_j V_{\alpha j}^*V_{\beta j}\left(\frac{E_j+p}{2E_j}e^{iE_jt}+\frac{E_j-p}{2E_j}e^{-iE_jt}\right) \nonumber \\
&=&\sum_j V_{\alpha j}^*V_{\beta j}\left(\frac{e^{iE_jt}+e^{-iE_jt}}{2}+\frac{p}{E_j}\frac{e^{iE_jt}-e^{-iE_jt}}{2}\right) \nonumber \\
&=&\sum_j V_{\alpha j}^*V_{\beta j}\left\{\cos (E_jt)+i\frac{p}{E_j}\sin (E_jt)\right\}, \\
A(\nu_{\alpha R}\to\nu_{\beta L})&=&\sum_j -V_{\alpha j}^*U_{\beta j}\frac{m_j}{2E_j}(e^{iE_jt}-e^{-iE_jt})
=\sum_j -iV_{\alpha j}^*U_{\beta j}\frac{m_j}{E_j}\sin (E_jt).
\end{eqnarray}
Then, the oscillation probabilities of right-handed neutrinos are calculated by squaring the absolute value of corresponding amplitude,
\begin{eqnarray}
P(\nu_{\alpha R}\to\nu_{\alpha R})&=&
1-\sum_j |V_{\alpha j}|^4\left\{\frac{m_j^2}{E_j^2}\sin^2 (E_jt)\right\}
\nonumber \\
&&\hspace{-1cm}-2\sum_{j<k}\left|V_{\alpha j}V_{\alpha k}\right|^2
\left\{2\sin^2 \left(\frac{\Delta E_{jk}t}{2}\right)+\frac{E_jE_k-p^2}{E_jE_k}\sin (E_jt)\sin (E_kt)\right\},
\label{r-pro1} \\
P(\nu_{\alpha R}\to\nu_{\beta R})&=&
-\sum_j |V_{\alpha j}V_{\beta j}|^2\left\{\frac{m_j^2}{E_j^2}\sin^2 (E_jt)\right\} \nonumber \\
&&\hspace{-1cm}-2\sum_{j<k}{\rm Re}\left[V_{\alpha j}V_{\beta j}^*V_{\alpha k}^*V_{\beta k}\right]
\left\{2\sin^2 \left(\frac{\Delta E_{jk}t}{2}\right)+\frac{E_jE_k-p^2}{E_jE_k}\sin (E_jt)\sin (E_kt)\right\} \nonumber \\
&&\hspace{-1cm}+2\sum_{j<k}{\rm Im}\left[V_{\alpha j}V_{\beta j}^*V_{\alpha k}^*V_{\beta k}\right]
\left\{\sin (\Delta E_{jk}t)+\frac{E_k-p}{E_k}\cos (E_jt)\sin (E_kt)-\frac{E_j-p}{E_j}\cos (E_kt)\sin (E_jt)\right\}, \\
P(\nu_{\alpha R}\to\nu_{\alpha L})&=&
\sum_j |V_{\alpha j}U_{\alpha j}|^2\frac{m_j^2}{E_j^2}\sin^2 (E_jt)
+2\sum_{j<k}{\rm Re}[V_{\alpha j}V_{\alpha k}^*U_{\alpha j}^*U_{\alpha k}]\frac{m_jm_k}{E_jE_k}\sin (E_jt)\sin (E_kt),\\
P(\nu_{\alpha R}\to\nu_{\beta L})&=&
\sum_j |V_{\alpha j}U_{\beta j}|^2\frac{m_j^2}{E_j^2}\sin^2 (E_jt)
+2\sum_{j<k}{\rm Re}[V_{\alpha j}V_{\alpha k}^*U_{\beta j}^*U_{\beta k}]\frac{m_jm_k}{E_jE_k}\sin (E_jt)\sin (E_kt).
\label{r-pro2}
\end{eqnarray}
These probabilities are obtained by exchanging the left-handed mixing matrix $U$ and the right-handed mixing matrix $V$
in (\ref{rela-survive})-(\ref{chirality-change2}).
There is a possibility that the new mixing angles for right-handed neutrinos could be measured
through $\nu_R \to \nu_R$ oscillations if the flavors of right-handed neutrino are distinguished.
In this case, high energy pion can decay through $W_R$ and produce $\nu_R$ beam.
If this $\nu_R$ reacts with some matter in a detector and puts it back to the right-handed charged lepton
through $W_R$, we can observe the charged lepton.
Thus, we measure the new mixing angles for right-handed neutrinos
without the suppression of order $(m/E)^2$.

The oscillation probabilities for $\nu_R^{\prime}$ are also obtained by changing the sign of the momentum $p$
in eqs.(\ref{r-pro1})-(\ref{r-pro2}) as
\begin{eqnarray}
P(\nu_{\alpha R}^{\prime}\to\nu_{\alpha R}^{\prime})&=&
1-\sum_j |V_{\alpha j}|^4\left\{\frac{m_j^2}{E_j^2}\sin^2 (E_jt)\right\} \nonumber \\
&&\hspace{-1cm}-2\sum_{j<k}\left|V_{\alpha j}V_{\alpha k}\right|^2
\left\{2\sin^2 \left(\frac{\Delta E_{jk}t}{2}\right)+\frac{E_jE_k-p^2}{E_jE_k}\sin (E_jt)\sin (E_kt)\right\},
\label{r-pro3} \\
P(\nu_{\alpha R}^{\prime}\to\nu_{\beta R}^{\prime})&=&
-\sum_j |V_{\alpha j}V_{\beta j}|^2\left\{\frac{m_j^2}{E_j^2}\sin^2 (E_jt)\right\}\nonumber \\
&&\hspace{-1cm}-2\sum_{j<k}{\rm Re}\left[V_{\alpha j}V_{\beta j}^*V_{\alpha k}^*V_{\beta k}\right]
\left\{2\sin^2 \left(\frac{\Delta E_{jk}t}{2}\right)+\frac{E_jE_k-p^2}{E_jE_k}\sin (E_jt)\sin (E_kt)\right\} \nonumber \\
&&\hspace{-1cm}-2\sum_{j<k}{\rm Im}\left[V_{\alpha j}V_{\beta j}^*V_{\alpha k}^*V_{\beta k}\right]
\left\{\sin (\Delta E_{jk}t)+\frac{E_k-p}{E_k}\cos (E_jt)\sin (E_kt)-\frac{E_j-p}{E_j}\cos (E_kt)\sin (E_jt)\right\}, \\
P(\nu_{\alpha R}^{\prime}\to\nu_{\alpha L}^{\prime})&=&
\sum_j |V_{\alpha j}U_{\alpha j}|^2\frac{m_j^2}{E_j^2}\sin^2 (E_jt)
+2\sum_{j<k}{\rm Re}[V_{\alpha j}V_{\alpha k}^*U_{\alpha j}^*U_{\alpha k}]\frac{m_jm_k}{E_jE_k}\sin (E_jt)\sin (E_kt),\\
P(\nu_{\alpha R}^{\prime}\to\nu_{\beta L}^{\prime})&=&
\sum_j |V_{\alpha j}U_{\beta j}|^2\frac{m_j^2}{E_j^2}\sin^2 (E_jt)
+2\sum_{j<k}{\rm Re}[V_{\alpha j}V_{\alpha k}^*U_{\beta j}^*U_{\beta k}]\frac{m_jm_k}{E_jE_k}\sin (E_jt)\sin (E_kt).
\label{r-pro4}
\end{eqnarray}
\end{widetext}
In the case of also right-handed neutrinos,
the sign of the term proportional to ${\rm Im}\left[V_{\alpha j}V_{\beta j}^*V_{\alpha k}^*V_{\beta k}\right]$
becomes opposite for $\nu$ and $\nu^{\prime}$.

%%%%%%%%%%%%%%%%%%%%%%%%%%%%%%
\subsection{Oscillation Probabilities of Anti-Neutrinos}

Next, we consider the oscillation probabilities for anti-neutrinos, 
which are defined as the charge conjugation of neutrinos.
The charge conjugation of $\psi_{\alpha}$,
\begin{eqnarray}
&&\psi_{\alpha}^c=(\psi_{\alpha L})^c+(\psi_{\alpha R})^c
\equiv \left(\begin{array}{c}\nu_{\alpha L}^c \\ \nu_{\alpha L}^{c\prime} \\
\nu_{\alpha R}^c \\ \nu_{\alpha R}^{c\prime} \end{array}\right) \nonumber \\
&&\equiv \left(\begin{array}{c}i\sigma_2 \eta_{\alpha}^* \\ -i\sigma_2 \xi_{\alpha}^{*} \end{array}\right)
=\left(\begin{array}{c}\nu_{\alpha L}^* \\ -\nu_{\alpha L}^{*\prime} \\ -\nu_{\alpha R}^* \\ \nu_{\alpha R}^{*\prime}, \end{array}\right)
\end{eqnarray}
also satisfies the Dirac equation,
\begin{eqnarray}
i\gamma^\mu \partial_\mu \psi_{\alpha L}^c-\sum_{\beta}m_{\beta\alpha} \psi_{\beta R}^c=0.
\end{eqnarray}
In general, the Dirac equation for $\psi^c$ is slightly different from that for $\psi$
because the mass term is complex in general. Namely, the mass term in the Dirac equation for $\psi^c$
becomes complex conjugate of that for $\psi$.
Multiplying $\gamma^0$ from the left, we obtain
\begin{eqnarray}
&&i\partial_0 \psi_{\alpha L}^c+i\gamma^0\gamma^i\partial_i \psi_{\alpha L}^c
-\sum_{\beta}m_{\beta\alpha} \gamma^0\psi_{\beta R}^c=0.
\end{eqnarray}
If we use two components spinors $\xi$ and $\eta$, the above equation can be rewritten as
\begin{eqnarray}
&&i\partial_0 \left(\!\!\begin{array}{c}i\sigma_2\eta_{\alpha}^* \\ 0\end{array}\!\!\right)
+i\left(\!\!\begin{array}{c}\sigma_i\partial_i (i\sigma_2\eta_{\alpha}^*) \\ 0\end{array}\!\!\right) \nonumber \\
&&-\sum_{\beta}m_{\beta\alpha} \left(\begin{array}{c}-i\sigma_2\xi_{\beta}^* \\ 0\end{array}\right)=0.
\end{eqnarray}
Taking out the upper two components, we obtain
\begin{eqnarray}
\hspace{-0.5cm}i\partial_0 (i\sigma_2\eta_{\alpha}^*)+i\sigma_i\partial_i (i\sigma_2\eta_{\alpha}^*)
-\sum_{\beta}m_{\beta\alpha} (-i\sigma_2\xi_{\beta}^*)=0.
\end{eqnarray}
In the same way, we also obtain
\begin{eqnarray}
\hspace{-0.5cm}i\partial_0 (-i\sigma_2\xi_{\alpha}^*)-i\sigma_i\partial_i (-i\sigma_2\xi_{\alpha}^*)
-\sum_{\beta}m_{\alpha\beta}^* (i\sigma_2\eta_{\beta}^*)=0.
\end{eqnarray}
Here, we take the complex conjugate of (\ref{equal-p}),
\begin{eqnarray}
\eta_{\alpha}^*(x,t)&=&e^{-i\vec{p}\cdot\vec{x}}\eta_{\alpha}^*(t)
=e^{-i\vec{p}\cdot\vec{x}}\left(\begin{array}{c}\nu_{\alpha L}^* \\ \nu_{\alpha L}^{*\prime}\end{array}\right),
\\
\xi_{\alpha}^*(x,t)&=&e^{-i\vec{p}\cdot\vec{x}}\xi_{\alpha}^*(t)
=e^{-i\vec{p}\cdot\vec{x}}\left(\begin{array}{c}\nu_{\alpha R}^* \\ \nu_{\alpha R}^{*\prime}\end{array}\right),
\label{equal-p-c}
\end{eqnarray}
and we choose $\vec{p}=(0,0,p)$. Then, the Dirac equations are rewritten as
\begin{eqnarray}
&&\hspace{-1.2cm}i\partial_0 \left(\begin{array}{c}\nu_{\alpha L}^{c} \\ \nu_{\alpha L}^{c\prime}\end{array}\right)
+p\left(\begin{array}{c}\nu_{\alpha L}^{c} \\ -\nu_{\alpha L}^{c\prime}\end{array}\right)
-\sum_{\beta}m_{\beta\alpha} \left(\begin{array}{c}\nu_{\beta R}^{c} \\ \nu_{\beta R}^{c\prime}\end{array}\right)=0, \\
&&\hspace{-1.2cm}i\partial_0 \left(\begin{array}{c}\nu_{\alpha R}^{c} \\ \nu_{\alpha R}^{c\prime}\end{array}\right)
-p\left(\begin{array}{c}\nu_{\alpha R}^{c} \\ -\nu_{\alpha R}^{c\prime}\end{array}\right)
-\sum_{\beta}m_{\alpha\beta}^* \left(\begin{array}{c}\nu_{\beta L}^{c} \\ \nu_{\beta L}^{c\prime}\end{array}\right)=0.
\end{eqnarray}
Combining the above equations for three flavors to one matrix form, the time evolution of the chirality-flavor eigenstates is given by
\begin{widetext}
\begin{eqnarray}
i\frac{d}{dt}\left(\begin{array}{c}
\nu_{eR}^{c\prime} \\ \nu_{\mu R}^{c\prime} \\ \nu_{\tau R}^{c\prime} \\ \nu_{eL}^{c\prime} \\ \nu_{\mu L}^{c\prime} \\ \nu_{\tau L}^{c\prime} \\
\nu_{eR}^{c} \\ \nu_{\mu R}^{c} \\ \nu_{\tau R}^{c} \\ \nu_{e L}^{c} \\ \nu_{\mu L}^{c} \\ \nu_{\tau L}^{c}
\end{array}\right)
=\left(\begin{array}{cccccc|cccccc}
-p & 0 & 0 & m_{ee}^* & m_{e\mu}^* & m_{e\tau}^* & 0 & 0 & 0 & 0 & 0 & 0 \\
0 & -p & 0 & m_{\mu e}^* & m_{\mu\mu}^* & m_{\mu\tau}^* & 0 & 0 & 0 & 0 & 0 & 0 \\
0 & 0 & -p & m_{\tau e}^* & m_{\tau\mu}^* & m_{\tau\tau}^* & 0 & 0 & 0 & 0 & 0 & 0 \\
m_{ee} & m_{\mu e} & m_{\tau e} & p & 0 & 0 & 0 & 0 & 0 & 0 & 0 & 0 \\
m_{e\mu} & m_{\mu\mu} & m_{\tau\mu} & 0 & p & 0 & 0 & 0 & 0 & 0 & 0 & 0 \\
m_{e\tau} & m_{\mu\tau} & m_{\tau\tau} & 0 & 0 & p & 0 & 0 & 0 & 0 & 0 & 0 \\
\hline
0 & 0 & 0 & 0 & 0 & 0 & p & 0 & 0 & m_{ee}^* & m_{e\mu}^* & m_{e\tau}^* \\
0 & 0 & 0 & 0 & 0 & 0 & 0 & p & 0 & m_{\mu e}^* & m_{\mu\mu}^* & m_{\mu\tau}^* \\
0 & 0 & 0 & 0 & 0 & 0 & 0 & 0 & p & m_{\tau e}^* & m_{\tau\mu}^* & m_{\tau\tau}^* \\
0 & 0 & 0 & 0 & 0 & 0 & m_{ee} & m_{\mu e} & m_{\tau e} & -p & 0 & 0 \\
0 & 0 & 0 & 0 & 0 & 0 & m_{e\mu} & m_{\mu\mu} & m_{\tau\mu} & 0 & -p & 0 \\
0 & 0 & 0 & 0 & 0 & 0 & m_{e\tau} & m_{\mu\tau} & m_{\tau\tau} & 0 & 0 & -p
\end{array}\right)
\left(\begin{array}{c}
\nu_{eR}^{c\prime} \\ \nu_{\mu R}^{c\prime} \\ \nu_{\tau R}^{c\prime} \\ \nu_{eL}^{c\prime} \\ \nu_{\mu L}^{c\prime} \\ \nu_{\tau L}^{c\prime} \\
\nu_{eR}^{c} \\ \nu_{\mu R}^{c} \\ \nu_{\tau R}^{c} \\ \nu_{e L}^{c} \\ \nu_{\mu L}^{c} \\ \nu_{\tau L}^{c}
\end{array}\right).
\end{eqnarray}
It is noted that this equation is for the anti-neutrinos with negative momentum $-p$ as seen from (\ref{equal-p-c}).
Changing the sign of momentum to calculate the oscillation probabilities of anti-neutrinos with positive momentum $p$,
\begin{eqnarray}
i\frac{d}{dt}\left(\begin{array}{c}
\nu_{eR}^{c\prime} \\ \nu_{\mu R}^{c\prime} \\ \nu_{\tau R}^{c\prime} \\ \nu_{eL}^{c\prime} \\ \nu_{\mu L}^{c\prime} \\ \nu_{\tau L}^{c\prime} \\
\nu_{eR}^{c} \\ \nu_{\mu R}^{c} \\ \nu_{\tau R}^{c} \\ \nu_{e L}^{c} \\ \nu_{\mu L}^{c} \\ \nu_{\tau L}^{c}
\end{array}\right)
=\left(\begin{array}{cccccc|cccccc}
p & 0 & 0 & m_{ee}^* & m_{e\mu}^* & m_{e\tau}^* & 0 & 0 & 0 & 0 & 0 & 0 \\
0 & p & 0 & m_{\mu e}^* & m_{\mu\mu}^* & m_{\mu\tau}^* & 0 & 0 & 0 & 0 & 0 & 0 \\
0 & 0 & p & m_{\tau e}^* & m_{\tau\mu}^* & m_{\tau\tau}^* & 0 & 0 & 0 & 0 & 0 & 0 \\
m_{ee} & m_{\mu e} & m_{\tau e} & -p & 0 & 0 & 0 & 0 & 0 & 0 & 0 & 0 \\
m_{e\mu} & m_{\mu\mu} & m_{\tau\mu} & 0 & -p & 0 & 0 & 0 & 0 & 0 & 0 & 0 \\
m_{e\tau} & m_{\mu\tau} & m_{\tau\tau} & 0 & 0 & -p & 0 & 0 & 0 & 0 & 0 & 0 \\
\hline
0 & 0 & 0 & 0 & 0 & 0 & -p & 0 & 0 & m_{ee}^* & m_{e\mu}^* & m_{e\tau}^* \\
0 & 0 & 0 & 0 & 0 & 0 & 0 & -p & 0 & m_{\mu e}^* & m_{\mu\mu}^* & m_{\mu\tau}^* \\
0 & 0 & 0 & 0 & 0 & 0 & 0 & 0 & -p & m_{\tau e}^* & m_{\tau\mu}^* & m_{\tau\tau}^* \\
0 & 0 & 0 & 0 & 0 & 0 & m_{ee} & m_{\mu e} & m_{\tau e} & p & 0 & 0 \\
0 & 0 & 0 & 0 & 0 & 0 & m_{e\mu} & m_{\mu\mu} & m_{\tau\mu} & 0 & p & 0 \\
0 & 0 & 0 & 0 & 0 & 0 & m_{e\tau} & m_{\mu\tau} & m_{\tau\tau} & 0 & 0 & p
\end{array}\right)
\left(\begin{array}{c}
\nu_{eR}^{c\prime} \\ \nu_{\mu R}^{c\prime} \\ \nu_{\tau R}^{c\prime} \\ \nu_{eL}^{c\prime} \\ \nu_{\mu L}^{c\prime} \\ \nu_{\tau L}^{c\prime} \\
\nu_{eR}^{c} \\ \nu_{\mu R}^{c} \\ \nu_{\tau R}^{c} \\ \nu_{e L}^{c} \\ \nu_{\mu L}^{c} \\ \nu_{\tau L}^{c}
\end{array}\right). \label{12-12-antinu-matrix}
\end{eqnarray}
Comparing this with (\ref{12-12-nu-matrix}), we can see that the replacements of
$\nu \to \nu^{c}$, $\nu^{\prime}\to \nu^{c\prime}$, $m\to m^*$ in (\ref{12-12-nu-matrix}) lead to
(\ref{12-12-antinu-matrix}).
According to this correspondence, we derive the oscillation probabilities for anti-neutrinos
by replacing $U \to U^*$ and $V\to V^*$ in (\ref{rela-survive})-(\ref{chirality-change2}).
Namely, we obtain the probabilities for anti-neutrinos with positive momentum $p$ as
\begin{eqnarray}
P(\nu_{\alpha L}^{c}\to\nu_{\alpha L}^{c})&=&
1-\sum_j |U_{\alpha j}|^4\left\{\frac{m_j^2}{E_j^2}\sin^2 (E_jt)\right\} \nonumber \\
&&\hspace{-1cm}-2\sum_{j<k}\left|U_{\alpha j}U_{\alpha k}\right|^2
\left\{2\sin^2 \left(\frac{\Delta E_{jk}t}{2}\right)+\frac{E_jE_k-p^2}{E_jE_k}\sin (E_jt)\sin (E_kt)\right\},
\label{naLc-naLc} \\
P(\nu_{\alpha L}^{c}\to\nu_{\beta L}^{c})&=&
-\sum_j |U_{\alpha j}U_{\beta j}|^2\left\{\frac{m_j^2}{E_j^2}\sin^2 (E_jt)\right\}\nonumber \\
&&\hspace{-1.5cm}-2\sum_{j<k}{\rm Re}\left[U_{\alpha j}U_{\beta j}^*U_{\alpha k}^*U_{\beta k}\right]
\left\{2\sin^2 \left(\frac{\Delta E_{jk}t}{2}\right)+\frac{E_jE_k-p^2}{E_jE_k}\sin (E_jt)\sin (E_kt)\right\} \nonumber \\
&&\hspace{-1.5cm}+2\sum_{j<k}{\rm Im}\left[U_{\alpha j}U_{\beta j}^*U_{\alpha k}^*U_{\beta k}\right]
\left\{\sin (\Delta E_{jk}t)+\frac{E_k-p}{E_k}\cos (E_jt)\sin (E_kt)-\frac{E_j-p}{E_j}\cos (E_kt)\sin (E_jt)\right\},
\label{rela-anti-nu-transition1} \\
P(\nu_{\alpha L}^{c}\to\nu_{\alpha R}^{c})&=&
\sum_j |U_{\alpha j}V_{\alpha j}|^2\frac{m_j^2}{E_j^2}\sin^2 (E_jt)
+2\sum_{j<k}{\rm Re}[U_{\alpha j}U_{\alpha k}^*V_{\alpha j}^*V_{\alpha k}]\frac{m_jm_k}{E_jE_k}\sin (E_jt)\sin (E_kt),\\
P(\nu_{\alpha L}^{c}\to\nu_{\beta R}^{c})&=&
\sum_j |U_{\alpha j}V_{\beta j}|^2\frac{m_j^2}{E_j^2}\sin^2 (E_jt)
+2\sum_{j<k}{\rm Re}[U_{\alpha j}U_{\alpha k}^*V_{\beta j}^*V_{\beta k}]\frac{m_jm_k}{E_jE_k}\sin (E_jt)\sin (E_kt),
\end{eqnarray}
and
\begin{eqnarray}
P(\nu_{\alpha L}^{c\prime}\to\nu_{\alpha L}^{c\prime})&=&
1-\sum_j |U_{\alpha j}|^4\left\{\frac{m_j^2}{E_j^2}\sin^2 (E_jt)\right\}\nonumber \\
&&\hspace{-1cm}-2\sum_{j<k}\left|U_{\alpha j}U_{\alpha k}\right|^2
\left\{2\sin^2 \left(\frac{\Delta E_{jk}t}{2}\right)+\frac{E_jE_k-p^2}{E_jE_k}\sin (E_jt)\sin (E_kt)\right\},
\\
P(\nu_{\alpha L}^{c\prime}\to\nu_{\beta L}^{c\prime})&=&
-\sum_j |U_{\alpha j}U_{\beta j}|^2\left\{\frac{m_j^2}{E_j^2}\sin^2 (E_jt)\right\}\nonumber \\
&&\hspace{-1.5cm}-2\sum_{j<k}{\rm Re}\left[U_{\alpha j}U_{\beta j}^*U_{\alpha k}^*U_{\beta k}\right]
\left\{2\sin^2 \left(\frac{\Delta E_{jk}t}{2}\right)+\frac{E_jE_k-p^2}{E_jE_k}\sin (E_jt)\sin (E_kt)\right\} \nonumber \\
&&\hspace{-1.5cm}-2\sum_{j<k}{\rm Im}\left[U_{\alpha j}U_{\beta j}^*U_{\alpha k}^*U_{\beta k}\right]
\left\{\sin (\Delta E_{jk}t)+\frac{E_k-p}{E_k}\cos (E_jt)\sin (E_kt)-\frac{E_j-p}{E_j}\cos (E_kt)\sin (E_jt)\right\},
\label{rela-anti-nu-transition2} 
\end{eqnarray}
\begin{eqnarray}
P(\nu_{\alpha L}^{c\prime}\to\nu_{\alpha R}^{c\prime})&=&
\sum_j |U_{\alpha j}V_{\alpha j}|^2\frac{m_j^2}{E_j^2}\sin^2 (E_jt)
+2\sum_{j<k}{\rm Re}[U_{\alpha j}U_{\alpha k}^*V_{\alpha j}^*V_{\alpha k}]\frac{m_jm_k}{E_jE_k}\sin (E_jt)\sin (E_kt),\\
P(\nu_{\alpha L}^{c\prime}\to\nu_{\beta R}^{c\prime})&=&
\sum_j |U_{\alpha j}V_{\beta j}|^2\frac{m_j^2}{E_j^2}\sin^2 (E_jt)
+2\sum_{j<k}{\rm Re}[U_{\alpha j}U_{\alpha k}^*V_{\beta j}^*V_{\beta k}]\frac{m_jm_k}{E_jE_k}\sin (E_jt)\sin (E_kt).
\label{naLcp-nbRcp}
\end{eqnarray}
Comparing (\ref{naLc-naLc})-(\ref{naLcp-nbRcp})
with (\ref{rela-survive})-(\ref{chirality-change2}) and (\ref{rela-survive-prime})-(\ref{naLp-nbRp}),
we can see that only the sign of the term (\ref{rela-anti-nu-transition1}) and (\ref{rela-anti-nu-transition2})
proportional to ${\rm Im}\left[U_{\alpha j}U_{\beta j}^*U_{\alpha k}^*U_{\beta k}\right]$
is different for neutrinos and anti-neutrinos.
The difference comes from the complex conjugate of $U$ and $V$ for anti-neutrinos.
It is noted that the probabilities with chirality-flip are the same for neutrinos and anti-neutrinos.

Furthermore, the oscillation probabilities for right-handed anti-neutrinos are given by the replacement,
$U \to U^*$ and $V\to V^*$ in (\ref{r-pro1})-(\ref{r-pro2}) and (\ref{r-pro3})-(\ref{r-pro4}),
\begin{eqnarray}
P(\nu_{\alpha R}^{c}\to\nu_{\alpha R}^{c})&=&
1-\sum_j |V_{\alpha j}|^4\left\{\frac{m_j^2}{E_j^2}\sin^2 (E_jt)\right\}\nonumber \\
&&\hspace{-1cm}-2\sum_{j<k}\left|V_{\alpha j}V_{\alpha k}\right|^2
\left\{2\sin^2 \left(\frac{\Delta E_{jk}t}{2}\right)+\frac{E_jE_k-p^2}{E_jE_k}\sin (E_jt)\sin (E_kt)\right\},
\\
P(\nu_{\alpha R}^{c}\to\nu_{\beta R}^{c})&=&
-\sum_j |V_{\alpha j}V_{\beta j}|^2\left\{\frac{m_j^2}{E_j^2}\sin^2 (E_jt)\right\}\nonumber \\
&&\hspace{-1.5cm}-2\sum_{j<k}{\rm Re}\left[V_{\alpha j}V_{\beta j}^*V_{\alpha k}^*V_{\beta k}\right]
\left\{2\sin^2 \left(\frac{\Delta E_{jk}t}{2}\right)+\frac{E_jE_k-p^2}{E_jE_k}\sin (E_jt)\sin (E_kt)\right\} \nonumber \\
&&\hspace{-1.5cm}-2\sum_{j<k}{\rm Im}\left[V_{\alpha j}V_{\beta j}^*V_{\alpha k}^*V_{\beta k}\right]
\left\{\sin (\Delta E_{jk}t)+\frac{E_k-p}{E_k}\cos (E_jt)\sin (E_kt)-\frac{E_j-p}{E_j}\cos (E_kt)\sin (E_jt)\right\}, \\
P(\nu_{\alpha R}^{c}\to\nu_{\alpha L}^{c})&=&
\sum_j |V_{\alpha j}U_{\alpha j}|^2\frac{m_j^2}{E_j^2}\sin^2 (E_jt)
+2\sum_{j<k}{\rm Re}[V_{\alpha j}V_{\alpha k}^*U_{\alpha j}^*U_{\alpha k}]\frac{m_jm_k}{E_jE_k}\sin (E_jt)\sin (E_kt),\\
P(\nu_{\alpha R}^{c}\to\nu_{\beta L}^{c})&=&
\sum_j |V_{\alpha j}U_{\beta j}|^2\frac{m_j^2}{E_j^2}\sin^2 (E_jt)
+2\sum_{j<k}{\rm Re}[V_{\alpha j}V_{\alpha k}^*U_{\beta j}^*U_{\beta k}]\frac{m_jm_k}{E_jE_k}\sin (E_jt)\sin (E_kt),
\end{eqnarray}
and
\begin{eqnarray}
P(\nu_{\alpha R}^{c\prime}\to\nu_{\alpha R}^{c\prime})&=&
1-\sum_j |V_{\alpha j}|^4\left\{\frac{m_j^2}{E_j^2}\sin^2 (E_jt)\right\}\nonumber \\
&&\hspace{-1cm}-2\sum_{j<k}\left|V_{\alpha j}V_{\alpha k}\right|^2
\left\{2\sin^2 \left(\frac{\Delta E_{jk}t}{2}\right)+\frac{E_jE_k-p^2}{E_jE_k}\sin (E_jt)\sin (E_kt)\right\},
\\
P(\nu_{\alpha R}^{c\prime}\to\nu_{\beta R}^{c\prime})&=&
-\sum_j |V_{\alpha j}V_{\beta j}|^2\left\{\frac{m_j^2}{E_j^2}\sin^2 (E_jt)\right\}\nonumber \\
&&\hspace{-1.5cm}-2\sum_{j<k}{\rm Re}\left[V_{\alpha j}V_{\beta j}^*V_{\alpha k}^*V_{\beta k}\right]
\left\{2\sin^2 \left(\frac{\Delta E_{jk}t}{2}\right)+\frac{E_jE_k-p^2}{E_jE_k}\sin (E_jt)\sin (E_kt)\right\} \nonumber \\
&&\hspace{-1.5cm}+2\sum_{j<k}{\rm Im}\left[V_{\alpha j}V_{\beta j}^*V_{\alpha k}^*V_{\beta k}\right]
\left\{\sin (\Delta E_{jk}t)+\frac{E_k-p}{E_k}\cos (E_jt)\sin (E_kt)-\frac{E_j-p}{E_j}\cos (E_kt)\sin (E_jt)\right\}, \\
P(\nu_{\alpha R}^{c\prime}\to\nu_{\alpha L}^{c\prime})&=&
\sum_j |V_{\alpha j}U_{\alpha j}|^2\frac{m_j^2}{E_j^2}\sin^2 (E_jt)
+2\sum_{j<k}{\rm Re}[V_{\alpha j}V_{\alpha k}^*U_{\alpha j}^*U_{\alpha k}]\frac{m_jm_k}{E_jE_k}\sin (E_jt)\sin (E_kt),\\
P(\nu_{\alpha R}^{c\prime}\to\nu_{\beta L}^{c\prime})&=&
\sum_j |V_{\alpha j}U_{\beta j}|^2\frac{m_j^2}{E_j^2}\sin^2 (E_jt)
+2\sum_{j<k}{\rm Re}[V_{\alpha j}V_{\alpha k}^*U_{\beta j}^*U_{\beta k}]\frac{m_jm_k}{E_jE_k}\sin (E_jt)\sin (E_kt).
\end{eqnarray}
\end{widetext}

%%%%%%%%%%%%%%%%%%%%%%%%%%%%%%%%%%%%%%%%%%%%%%%%%%%%%%%%%%
\section{Oscillation Probabilities of Majorana Neutrino From Relativistic Equation}

In this section, we derive the oscillation probabilities of the Majorana neutrinos in three generations or more.
We count the number of measurable mixing angles and CP phases in the case of $n$-generations.
In the non-relativistic method, there was no difference between the probabilities of the Dirac neutrinos and
the Majorana neutrinos.
However, there appears a difference in the oscillations with chirality-flip when we use the relativistic equation. \\

\subsection{Oscillation Probabilities of Neutrinos}

The lagrangian for Majorana neutrinos in three generations is given by
\begin{eqnarray}
&&L=\displaystyle{\sum_{\alpha}\frac{1}{2}\left[i\overline{\psi_{\alpha L}}\gamma^\mu \partial_\mu \psi_{\alpha L}
+i\overline{\psi_{\alpha L}^c}\gamma^\mu \partial_\mu \psi_{\alpha L}^c\right]} \nonumber \\
&&-\displaystyle{\sum_{(\alpha, \beta)}\frac{1}{2}\left[
\overline{\psi_{\beta L}^c}M_{\beta\alpha}\psi_{\alpha L}
-\overline{\psi_{\alpha L}}M_{\beta\alpha}^*\psi_{\beta L}^c
\right]}.
\end{eqnarray}
About the kinetic term in the lagrangian, the relation,
\begin{eqnarray}
{\cal L}_{\rm kin}=i\overline{\psi_{\alpha L}}\gamma^\mu \partial_\mu \psi_{\alpha L}
=i\overline{\psi_{\alpha L}^c}\gamma^\mu \partial_\mu \psi_{\alpha L}^c,
\end{eqnarray}
holds and the Eular-Lagrange equation for $\overline{\psi_{\alpha L}}$,
\begin{eqnarray}
\frac{\partial L}{\partial \overline{\psi_{\alpha L}}}
-\partial_\mu \left(\frac{\partial L}{\partial(\partial_\mu \overline{\psi_{\alpha L}})}\right)=0,
\end{eqnarray}
leads to
\begin{eqnarray}
i\gamma^\mu \partial_\mu \psi_{\alpha L}-\sum_{\beta}M_{\beta\alpha}^* \psi_{\beta L}^c=0.
\end{eqnarray}
Multiplying $\gamma_0$ from the left, this equation can be rewritten as
\begin{eqnarray}
i\partial_0 \psi_{\alpha L}+i\gamma^0\gamma^i\partial_i \psi_{\alpha L}
-\sum_{\beta}M_{\beta\alpha}^* \gamma^0\psi_{\beta L}^c=0.
\end{eqnarray}
Substituting (\ref{gamma-mat}) and (\ref{psi-def}) into this equation,
we obtain the equation for two-component spinor $\eta$,
\begin{eqnarray}
&&i\partial_0 \left(\begin{array}{c}0 \\ \eta_{\alpha}\end{array}\right)+i\left(\begin{array}{cc}0 & 1 \\ 1 & 0\end{array}\right)
\left(\begin{array}{cc}0 & -\sigma_i \\ \sigma_i & 0\end{array}\right)\partial_i
\left(\begin{array}{c}0 \\ \eta_{\alpha}\end{array}\right) \nonumber \\
&&-\sum_{\beta}M_{\beta\alpha}^* \left(\begin{array}{cc}0 & 1 \\ 1 & 0\end{array}\right)\left(\begin{array}{c}i\sigma_2\eta_{\beta}^* \\ 0\end{array}\right)=0, \\
&&\hspace{-1cm}i\partial_0 \left(\begin{array}{c}0 \\ \eta_{\alpha}\end{array}\right)-i
\left(\!\!\begin{array}{c}0 \\ \sigma_i\partial_i\eta_{\alpha}\end{array}\!\!\right)
-\sum_{\beta}M_{\beta\alpha}^* \left(\!\!\begin{array}{c}0 \\ i\sigma_2\eta_{\beta}^*\end{array}\!\!\right)=0.
\end{eqnarray}
Taking out the lower two components, we obtain
\begin{eqnarray}
&&i\partial_0 \eta_{\alpha}-i\sigma_i\partial_i \eta_{\alpha}
-\sum_{\beta}M_{\beta\alpha}^* (i\sigma_2\eta_{\beta}^*)=0. \label{eom-eta-a}
\end{eqnarray}
In the same way, the Eular-Lagrange equation for $\overline{\psi_{\alpha L}^c}$,
\begin{eqnarray}
\frac{\partial L}{\partial \overline{\psi_{\alpha L}^c}}
-\partial_\mu \left(\frac{\partial L}{\partial(\partial_\mu \overline{\psi_{\alpha L}^c})}\right)=0,
\end{eqnarray}
leads to the equation,
\begin{eqnarray}
i\gamma^\mu \partial_\mu \psi_{\alpha L}^c-\sum_{\beta}M_{\alpha\beta} \psi_{\beta L}=0.
\end{eqnarray}
Multiplying $\gamma_0$ from the left, the above equation becomes
\begin{eqnarray}
&&i\partial_0 \psi_{\alpha L}^c+i\gamma^0\gamma^i\partial_i \psi_{\alpha L}^c
-\sum_{\beta}M_{\alpha\beta} \gamma^0\psi_{\beta L}=0.
\end{eqnarray}
Substituting (\ref{gamma-mat}) and (\ref{psi-def}) into this equation, we obtain the equation for two-component spinor $\eta$,
\begin{eqnarray}
&&i\partial_0 \left(\begin{array}{c}i\sigma_2\eta_{\alpha}^* \\ 0\end{array}\right)
+i\left(\begin{array}{cc}0 & 1 \\ 1 & 0\end{array}\right)
\left(\begin{array}{cc}0 & -\sigma_i \\ \sigma_i & 0\end{array}\right)\partial_i
\left(\begin{array}{c}i\sigma_2\eta_{\alpha}^* \\ 0\end{array}\right) \nonumber \\
&&-\sum_{\beta}M_{\alpha\beta} \left(\begin{array}{cc}0 & 1 \\ 1 & 0\end{array}\right)\left(\begin{array}{c}0 \\ \eta_{\beta}\end{array}\right)=0, \\
&&\hspace{-0.5cm}i\partial_0 \left(\!\!\begin{array}{c}i\sigma_2\eta_{\alpha}^* \\ 0\end{array}\!\!\right)\!\!+i
\left(\!\!\begin{array}{c}\sigma_i\partial_i (i\sigma_2\eta_{\alpha}^*) \\ 0\end{array}\!\!\right)
\!\!-\sum_{\beta}M_{\alpha\beta} \left(\!\!\begin{array}{c}\eta_{\beta} \\ 0\end{array}\!\!\right)=0.
\end{eqnarray}
Taking out the upper two components, we obtain
\begin{eqnarray}
&&i\partial_0 (i\sigma_2\eta_{\alpha}^*)+i\sigma_i\partial_i (i\sigma_2\eta_{\alpha}^*)
-\sum_{\beta}M_{\alpha\beta} \eta_{\beta}=0. \label{eom-etac-a}
\end{eqnarray}
Here, we take the equal momentum assumption for all flavors,
\begin{eqnarray}
\eta_{\alpha}(x,t)=e^{i\vec{p}\cdot\vec{x}}\eta_{\alpha}(t)
=e^{i\vec{p}\cdot\vec{x}}\left(\begin{array}{c}\nu_{\alpha L}^{\prime} \\ \nu_{\alpha L}\end{array}\right), \label{eta}
\end{eqnarray}
and we choose $\vec{p}=(0,0,p)$.
The complex conjugate of these two-component spinors is given by
\begin{eqnarray}
\eta_{\alpha}^*(x,t)=e^{-i\vec{p}\cdot\vec{x}}\eta_{\alpha}^*(t)
=e^{-i\vec{p}\cdot\vec{x}}\left(\begin{array}{c}\nu_{\alpha L}^{*\prime} \\ \nu_{\alpha L}^*\end{array}\right). \label{etastar}
\end{eqnarray}
It is noted that $\nu^*$ included in $\eta^*$ has the negative momentum.
Substituting (\ref{eta}) and (\ref{etastar}) into (\ref{eom-eta-a}) and (\ref{eom-etac-a}),
the time evolution of the chirality-flavor eigenstates is given by
\begin{widetext}
\begin{eqnarray}
i\frac{d}{dt}\left(\begin{array}{c}
\nu_{eL}^{c\prime} \\ \nu_{\mu L}^{c\prime} \\ \nu_{\tau L}^{c\prime} \\ \nu_{eL}^{\prime} \\ \nu_{\mu L}^{\prime} \\ \nu_{\tau L}^{\prime} \\
\nu_{eL}^c \\ \nu_{\mu L}^c \\ \nu_{\tau L}^c \\ \nu_{e L} \\ \nu_{\mu L} \\ \nu_{\tau L}
\end{array}\right)
=\left(\begin{array}{cccccc|cccccc}
p & 0 & 0 & M_{ee} & M_{e\mu} & M_{e\tau} & 0 & 0 & 0 & 0 & 0 & 0 \\
0 & p & 0 & M_{\mu e} & M_{\mu\mu} & M_{\mu\tau} & 0 & 0 & 0 & 0 & 0 & 0 \\
0 & 0 & p & M_{\tau e} & M_{\tau\mu} & M_{\tau\tau} & 0 & 0 & 0 & 0 & 0 & 0 \\
M_{ee}^* & M_{\mu e}^* & M_{\tau e}^* & -p & 0 & 0 & 0 & 0 & 0 & 0 & 0 & 0 \\
M_{e\mu}^* & M_{\mu\mu}^* & M_{\tau\mu}^* & 0 & -p & 0 & 0 & 0 & 0 & 0 & 0 & 0 \\
M_{e\tau}^* & M_{\mu\tau}^* & M_{\tau\tau}^* & 0 & 0 & -p & 0 & 0 & 0 & 0 & 0 & 0 \\
\hline
0 & 0 & 0 & 0 & 0 & 0 & -p & 0 & 0 & M_{ee} & M_{e\mu} & M_{e\tau} \\
0 & 0 & 0 & 0 & 0 & 0 & 0 & -p & 0 & M_{\mu e} & M_{\mu\mu} & M_{\mu\tau} \\
0 & 0 & 0 & 0 & 0 & 0 & 0 & 0 & -p & M_{\tau e} & M_{\tau\mu} & M_{\tau\tau} \\
0 & 0 & 0 & 0 & 0 & 0 & M_{ee}^* & M_{\mu e}^* & M_{\tau e}^* & p & 0 & 0 \\
0 & 0 & 0 & 0 & 0 & 0 & M_{e\mu}^* & M_{\mu\mu}^* & M_{\tau\mu}^* & 0 & p & 0 \\
0 & 0 & 0 & 0 & 0 & 0 & M_{e\tau}^* & M_{\mu\tau}^* & M_{\tau\tau}^* & 0 & 0 & p
\end{array}\right)
\left(\begin{array}{c}
\nu_{eL}^{c\prime} \\ \nu_{\mu L}^{c\prime} \\ \nu_{\tau L}^{c\prime} \\ \nu_{eL}^{\prime} \\ \nu_{\mu L}^{\prime} \\ \nu_{\tau L}^{\prime} \\
\nu_{eL}^c \\ \nu_{\mu L}^c \\ \nu_{\tau L}^c \\ \nu_{e L} \\ \nu_{\mu L} \\ \nu_{\tau L}
\end{array}\right), \label{12-12-nu-matrix-majo}
\end{eqnarray}
as in the case of two generations.
In the case of the Dirac neutrinos,
neutrinos and anti-neutrinos belong to different multiplet and do not mix each other,
accordingly, the lepton number conservation is maintained.
On the other hand, in the case of the Majorana neutrinos, neutrinos and anti-neutrinos are in the same
multiplet by the existence of the Majorana mass term and neutrinos can oscillate to anti-neutrinos.
We would like to emphasize that neutrinos and anti-neutrinos have opposite momentum.
This is inevitable in the case of the Majorana neutrinos because both $\eta$ and $\eta^*$ are included
in an equation, unlike the Dirac neutrinos.
It is also noted that $\nu$, $\nu^c$ and $\nu^{\prime}$, $\nu^{c\prime}$ are separated completely and cannot be mixed.
The Dirac equation for $\nu$ and $\nu^c$ is given by
\begin{eqnarray}
i\frac{d}{dt}\left(\begin{array}{c}
\nu_{eL}^{c} \\ \nu_{\mu L}^{c} \\ \nu_{\tau L}^{c} \\ \nu_{e L} \\ \nu_{\mu L} \\ \nu_{\tau L}
\end{array}\right)
=\left(\begin{array}{ccc|ccc}
-p & 0 & 0 & M_{ee} & M_{e\mu} & M_{e\tau} \\
0 & -p & 0 & M_{\mu e} & M_{\mu\mu} & M_{\mu\tau} \\
0 & 0 & -p & M_{\tau e} & M_{\tau\mu} & M_{\tau\tau} \\
\hline
M_{ee}^* & M_{\mu e}^* & M_{\tau e}^* & p & 0 & 0 \\
M_{e\mu}^* & M_{\mu\mu}^* & M_{\tau\mu}^* & 0 & p & 0 \\
M_{e\tau}^* & M_{\mu\tau}^* & M_{\tau\tau}^* & 0 & 0 & p
\end{array}\right)
\left(\begin{array}{c}
\nu_{eL}^{c} \\ \nu_{\mu L}^{c} \\ \nu_{\tau L}^{c} \\ \nu_{e L} \\ \nu_{\mu L} \\ \nu_{\tau L}
\end{array}\right). \label{time-evolution-6-6}
\end{eqnarray}
This has the same constructure as the equation (\ref{Dirac-Hamiltonian}) in the previous section.
The only different point is that the mass matrix in the Hamiltonian is complex symmetric.
Therefore, the mass matrix can be diagonalized by one unitary matrix $U$.
Then, the chirality-flavor eigenstates are represented as the linear combination of
the chirality-mass eigenstates as
\begin{eqnarray}
\left(\begin{array}{c}
\nu_{eL}^{c} \\ \nu_{\mu L}^{c} \\ \nu_{\tau L}^{c} \\ \nu_{e L} \\ \nu_{\mu L} \\ \nu_{\tau L}
\end{array}\right)
=\left(\begin{array}{ccc|ccc}
U_{e1}^* & U_{e2}^* & U_{e3}^* & 0 & 0 & 0 \\
U_{\mu 1}^* & U_{\mu 2}^* & U_{\mu 3}^* & 0 & 0 & 0 \\
U_{\tau 1}^* & U_{\tau 2}^* & U_{\tau 3}^* & 0 & 0 & 0 \\
\hline
0 & 0 & 0 & U_{e1} & U_{e2} & U_{e3} \\
0 & 0 & 0 & U_{\mu 1} & U_{\mu 2} & U_{\mu 3} \\
0 & 0 & 0 & U_{\tau 1} & U_{\tau 2} & U_{\tau 3}
\end{array}\right)\left(\begin{array}{c}
\nu_{1L}^{c} \\ \nu_{2L}^{c} \\ \nu_{3L}^{c} \\ \nu_{1L} \\ \nu_{2L} \\ \nu_{3L}
\end{array}\right).
\end{eqnarray}
Namely, we obtain the oscillation probabilities for Majorana neutrino by the replacement
$V\to U^*$, $\nu_{\alpha R} \to \nu_{\alpha L}^{c}$ in eqs.(\ref{rela-survive})-(\ref{chirality-change2}),
\begin{eqnarray}
P(\nu_{\alpha L}\to\nu_{\alpha L})&=&
1-\sum_j |U_{\alpha j}|^4\left\{\frac{m_j^2}{E_j^2}\sin^2 (E_jt)\right\} \nonumber \\
&&\hspace{-1.5cm}-2\sum_{j<k}\left|U_{\alpha j}U_{\alpha k}\right|^2
\left\{2\sin^2 \left(\frac{\Delta E_{jk}t}{2}\right)+\frac{E_jE_k-p^2}{E_jE_k}\sin (E_jt)\sin (E_kt)\right\}, \label{aL-aL} 
\end{eqnarray}
\begin{eqnarray}
P(\nu_{\alpha L}\to\nu_{\beta L})&=&
-\sum_j |U_{\alpha j}U_{\beta j}|^2\left\{\frac{m_j^2}{E_j^2}\sin^2 (E_jt)\right\}\nonumber \\
&&\hspace{-1.5cm}-2\sum_{j<k}{\rm Re}\left[U_{\alpha j}U_{\beta j}^*U_{\alpha k}^*U_{\beta k}\right]
\left\{2\sin^2 \left(\frac{\Delta E_{jk}t}{2}\right)+\frac{E_jE_k-p^2}{E_jE_k}\sin (E_jt)\sin (E_kt)\right\} \nonumber \\
&&\hspace{-1.5cm}-2\sum_{j<k}{\rm Im}\left[U_{\alpha j}U_{\beta j}^*U_{\alpha k}^*U_{\beta k}\right]
\left\{\sin (\Delta E_{jk}t)+\frac{E_k-p}{E_k}\cos (E_jt)\sin (E_kt)-\frac{E_j-p}{E_j}\cos (E_kt)\sin (E_jt)\right\}, \\
P(\nu_{\alpha L}\to\nu_{\alpha L}^{c})&=&
\sum_j |U_{\alpha j}U_{\alpha j}|^2\frac{m_j^2}{E_j^2}\sin^2 (E_jt)
+2\sum_{j<k}{\rm Re}[U_{\alpha j}U_{\alpha k}^*U_{\alpha j}U_{\alpha k}^*]\frac{m_jm_k}{E_jE_k}\sin (E_jt)\sin (E_kt),
\label{aL-aLc}\\
P(\nu_{\alpha L}\to\nu_{\beta L}^{c})&=&
\sum_j |U_{\alpha j}U_{\beta j}|^2\frac{m_j^2}{E_j^2}\sin^2 (E_jt)
+2\sum_{j<k}{\rm Re}[U_{\alpha j}U_{\alpha k}^*U_{\beta j}U_{\beta k}^*]\frac{m_jm_k}{E_jE_k}\sin (E_jt)\sin (E_kt).
\label{aL-bLc}
\end{eqnarray}
In this way, we can derive the oscillation probabilites for neutrinos and anti-neutrinos
by entering $\nu_L^c$ in the same multiplet as $\nu_L$ instead of $\nu_R$.
The probabilities obtained above are described by the parameter independent manner.

Next, let us derive the probabilities for $\nu^{\prime}$ and $\nu^{c\prime}$.
The top-left part of the Hamiltonian in (\ref{12-12-nu-matrix-majo}) is the same structure as
the Hamiltonian in (\ref{Dirac-Hamiltonian2}).
We only have to replace $V\to U^*$ in (\ref{rela-survive-prime})-(\ref{naLp-nbRp}) according to the replacement $m\to M$ in the Hamiltonian.
Therefore, the oscillation probabilities for $\nu^{\prime}$ are given by
\begin{eqnarray}
P(\nu_{\alpha L}^{\prime}\to\nu_{\alpha L}^{\prime})&=&
1-\sum_j |U_{\alpha j}|^4\left\{\frac{m_j^2}{E_j^2}\sin^2 (E_jt)\right\} \nonumber \\
&&\hspace{-1.5cm}-2\sum_{j<k}\left|U_{\alpha j}U_{\alpha k}\right|^2
\left\{2\sin^2 \left(\frac{\Delta E_{jk}t}{2}\right)+\frac{E_jE_k-p^2}{E_jE_k}\sin (E_jt)\sin (E_kt)\right\}, \label{aLp-aLp}\\
P(\nu_{\alpha L}^{\prime}\to\nu_{\beta L}^{\prime})&=&
-\sum_j |U_{\alpha j}U_{\beta j}|^2\left\{\frac{m_j^2}{E_j^2}\sin^2 (E_jt)\right\}\nonumber \\
&&\hspace{-1.5cm}-2\sum_{j<k}{\rm Re}\left[U_{\alpha j}U_{\beta j}^*U_{\alpha k}^*U_{\beta k}\right]
\left\{2\sin^2 \left(\frac{\Delta E_{jk}t}{2}\right)+\frac{E_jE_k-p^2}{E_jE_k}\sin (E_jt)\sin (E_kt)\right\} \nonumber \\
&&\hspace{-1.5cm}+2\sum_{j<k}{\rm Im}\left[U_{\alpha j}U_{\beta j}^*U_{\alpha k}^*U_{\beta k}\right]
\left\{\sin (\Delta E_{jk}t)+\frac{E_k-p}{E_k}\cos (E_jt)\sin (E_kt)-\frac{E_j-p}{E_j}\cos (E_kt)\sin (E_jt)\right\},
\label{mnaLp-mnaLpi}\\
P(\nu_{\alpha L}^{\prime}\to\nu_{\alpha L}^{c\prime})&=&
\sum_j |U_{\alpha j}U_{\alpha j}|^2\frac{m_j^2}{E_j^2}\sin^2 (E_jt)
+2\sum_{j<k}{\rm Re}[U_{\alpha j}U_{\alpha k}^*U_{\alpha j}U_{\alpha k}^*]\frac{m_jm_k}{E_jE_k}\sin (E_jt)\sin (E_kt),
\label{aLp-aLcp}\\
P(\nu_{\alpha L}^{\prime}\to\nu_{\beta L}^{c\prime})&=&
\sum_j |U_{\alpha j}U_{\beta j}|^2\frac{m_j^2}{E_j^2}\sin^2 (E_jt)
+2\sum_{j<k}{\rm Re}[U_{\alpha j}U_{\alpha k}^*U_{\beta j}U_{\beta k}^*]\frac{m_jm_k}{E_jE_k}\sin (E_jt)\sin (E_kt).
\label{aLp-bLcp}
\end{eqnarray}
\end{widetext}
The difference between the probabilities for $\nu$ and $\nu^{\prime}$ appears in (\ref{mnaLp-mnaLpi}).
Namely, the sign of the term proportional to ${\rm Im}\left[U_{\alpha j}U_{\beta j}^*U_{\alpha k}^*U_{\beta k}\right]$
is reversed.
It is noted that eqs.(\ref{aL-aLc}), (\ref{aL-bLc}), (\ref{aLp-aLcp}) and (\ref{aLp-bLcp})
are the probabilities for oscillations from neutrinos with positive momentum to anti-neutrinos with negative momentum.

The probabilities of the Majorana neutrinos are obtained by
the replacement $\phi_{jL} \rightarrow \phi_j$, $\phi_{jR}\rightarrow -\phi_j$ and $V \rightarrow U^*$
in those of the Dirac neutrinos.
The probabilities without chirality-flip are the same as those of the Dirac neutrinos.
The probabilities with chirality-flip (\ref{aL-aLc}) and (\ref{aL-bLc}) depend on the Majorana CP phases through
the real part of the product of four matrix elements,
\begin{eqnarray}
&&\hspace{-0.5cm}{\rm Re}[U_{\alpha j}U_{\alpha k}^*U_{\beta j}U_{\beta k}^*] \nonumber \\
&&\hspace{-0.5cm}={\rm Re}[e^{i\rho_\alpha}\tilde{U}_{\alpha j}e^{i\phi_j}
e^{-i\rho_\alpha}\tilde{U}_{\alpha k}^*e^{-i\phi_k}
e^{i\rho_\beta}\tilde{U}_{\beta j}e^{i\phi_j}e^{-i\rho_\beta}\tilde{U}_{\beta k}^*
e^{-i\phi_k}] \nonumber \\
&&\hspace{-0.5cm}={\rm Re}[\tilde{U}_{\alpha j}\tilde{U}_{\alpha k}^*
\tilde{U}_{\beta j}\tilde{U}_{\beta k}^*
e^{2i(\phi_j-\phi_k)}] \nonumber \\
&&\hspace{-0.5cm}={\rm Re}(\tilde{U}_{\alpha j}\tilde{U}_{\alpha k}^*
\tilde{U}_{\beta j}\tilde{U}_{\beta k}^*)
\cos \{2(\phi_j-\phi_k)\} \nonumber \\
&&-{\rm Im}(\tilde{U}_{\alpha j}\tilde{U}_{\alpha k}^*
\tilde{U}_{\beta j}\tilde{U}_{\beta k}^*)
\sin \{2(\phi_j-\phi_k)\}.
\end{eqnarray}
As in the case of the Dirac neutrinos, we can obtain the information from both the sine and the cosine term
though there is no direct CP violation.
If we define $\Delta \phi_{jk}=\phi_j-\phi_k$, the probabilities of the Majorana neutrinos
depend on the new CP phase through the form $\Delta \phi_{jk}$.
As there are the relations like $\Delta \phi_{13}=\Delta \phi_{12}-\Delta \phi_{23}$,
independent parameters related to the new CP phases is two in three generations.
Namely, the number of CP phase appeared in the Majorana case is the same as that in the Dirac case.
The result obtained here is the same as previously known in the case of the Majorana neutrinos.

%%%%%%%%%%%%%%%%%%%%%%%%%%%
\begin{widetext}
\subsection{Oscillation Probabilities of Anti-Neutrinos}

Next, we consider the oscillation probabilities of anti-neutrinos with positive momentum.
As the anti-neutrinos have negative momentum in eq.(\ref{time-evolution-6-6}),
let us change the sign of momentum $p$ in order to derive the probabilities of anti-neutrinos with positive momentum.
Namely, we start from the time evolution equation
%\begin{widetext}
\begin{eqnarray}
i\frac{d}{dt}\left(\begin{array}{c}
\nu_{eL}^{c} \\ \nu_{\mu L}^{c} \\ \nu_{\tau L}^{c} \\ \nu_{e L} \\ \nu_{\mu L} \\ \nu_{\tau L}
\end{array}\right)
=\left(\begin{array}{ccc|ccc}
p & 0 & 0 & M_{ee} & M_{e\mu} & M_{e\tau} \\
0 & p & 0 & M_{\mu e} & M_{\mu\mu} & M_{\mu\tau} \\
0 & 0 & p & M_{\tau e} & M_{\tau\mu} & M_{\tau\tau} \\
\hline
M_{ee}^* & M_{\mu e}^* & M_{\tau e}^* & -p & 0 & 0 \\
M_{e\mu}^* & M_{\mu\mu}^* & M_{\tau\mu}^* & 0 & -p & 0 \\
M_{e\tau}^* & M_{\mu\tau}^* & M_{\tau\tau}^* & 0 & 0 & -p
\end{array}\right)
\left(\begin{array}{c}
\nu_{eL}^{c} \\ \nu_{\mu L}^{c} \\ \nu_{\tau L}^{c} \\ \nu_{e L} \\ \nu_{\mu L} \\ \nu_{\tau L}
\end{array}\right),
\end{eqnarray}
where anti-neutrinos $\nu^c$ have positive momentum and neutrinos $\nu$ have negative momentum.
Exchanging some rows and some columns, and using the symmetry of the Majorana mass term, this equation can be rewritten as
\begin{eqnarray}
i\frac{d}{dt}\left(\begin{array}{c}
\nu_{e L} \\ \nu_{\mu L} \\ \nu_{\tau L} \\ \nu_{eL}^{c} \\ \nu_{\mu L}^{c} \\ \nu_{\tau L}^{c}
\end{array}\right)
=\left(\begin{array}{ccc|ccc}
-p & 0 & 0 & M_{ee}^* & M_{e\mu}^* & M_{e\tau}^* \\
0 & -p & 0 & M_{\mu e}^* & M_{\mu\mu}^* & M_{\mu\tau}^* \\
0 & 0 & -p & M_{\tau e}^* & M_{\tau\mu}^* & M_{\tau\tau}^* \\
\hline
M_{ee} & M_{\mu e} & M_{\tau e} & p & 0 & 0 \\
M_{e\mu} & M_{\mu\mu} & M_{\tau\mu} & 0 & p & 0 \\
M_{e\tau} & M_{\mu\tau} & M_{\tau\tau} & 0 & 0 & p
\end{array}\right)
\left(\begin{array}{c}
\nu_{e L} \\ \nu_{\mu L} \\ \nu_{\tau L} \\ \nu_{eL}^{c} \\ \nu_{\mu L}^{c} \\ \nu_{\tau L}^{c}
\end{array}\right).
\end{eqnarray}
We can see that this equation is obtained by the exchange $M \leftrightarrow M^*$ and
$\nu \leftrightarrow \nu^c$ in eq.(\ref{time-evolution-6-6}).
Therefore, the probabilities of anti-neutrinos are obtained by the exchange $\nu \leftrightarrow \nu^c$ and
$U \leftrightarrow U^*$ in (\ref{aL-aL})-(\ref{aL-bLc}),
\begin{eqnarray}
P(\nu_{\alpha L}^c\to\nu_{\alpha L}^c)&=&
1-\sum_j |U_{\alpha j}|^4\left\{\frac{m_j^2}{E_j^2}\sin^2 (E_jt)\right\}\nonumber \\
&&\hspace{-1cm}-2\sum_{j<k}\left|U_{\alpha j}U_{\alpha k}\right|^2
\left\{2\sin^2 \left(\frac{\Delta E_{jk}t}{2}\right)+\frac{E_jE_k-p^2}{E_jE_k}\sin (E_jt)\sin (E_kt)\right\},
\\
P(\nu_{\alpha L}^c\to\nu_{\beta L}^c)&=&
-\sum_j |U_{\alpha j}U_{\beta j}|^2\left\{\frac{m_j^2}{E_j^2}\sin^2 (E_jt)\right\}\nonumber \\
&&\hspace{-1.5cm}-2\sum_{j<k}{\rm Re}\left[U_{\alpha j}^*U_{\beta j}U_{\alpha k}U_{\beta k}^*\right]
\left\{2\sin^2 \left(\frac{\Delta E_{jk}t}{2}\right)+\frac{E_jE_k-p^2}{E_jE_k}\sin (E_jt)\sin (E_kt)\right\} \nonumber \\
&&\hspace{-1.5cm}+2\sum_{j<k}{\rm Im}\left[U_{\alpha j}^*U_{\beta j}U_{\alpha k}U_{\beta k}^*\right]
\left\{\sin (\Delta E_{jk}t)+\frac{E_k-p}{E_k}\cos (E_jt)\sin (E_kt)-\frac{E_j-p}{E_j}\cos (E_kt)\sin (E_jt)\right\},
\label{aLc-bLc}\\
P(\nu_{\alpha L}^c\to\nu_{\alpha L})&=&
\sum_j |U_{\alpha j}U_{\alpha j}|^2\frac{m_j^2}{E_j^2}\sin^2 (E_jt)
+2\sum_{j<k}{\rm Re}[U_{\alpha j}U_{\alpha k}^*U_{\alpha j}U_{\alpha k}^*]\frac{m_jm_k}{E_jE_k}\sin (E_jt)\sin (E_kt),\\
P(\nu_{\alpha L}^c\to\nu_{\beta L})&=&
\sum_j |U_{\alpha j}U_{\beta j}|^2\frac{m_j^2}{E_j^2}\sin^2 (E_jt)
+2\sum_{j<k}{\rm Re}[U_{\alpha j}U_{\alpha k}^*U_{\beta j}U_{\beta k}^*]\frac{m_jm_k}{E_jE_k}\sin (E_jt)\sin (E_kt).
\end{eqnarray}
Comparing these probabilities with (\ref{aL-aL})-(\ref{aL-bLc}),
it turns out that only the sign in (\ref{aLc-bLc}) is reversed.
The oscillation probabilities for $\nu^{c\prime}$ are also obtained as
\begin{eqnarray}
P(\nu_{\alpha L}^{c\prime}\to\nu_{\alpha L}^{c\prime})&=&
1-\sum_j |U_{\alpha j}|^4\left\{\frac{m_j^2}{E_j^2}\sin^2 (E_jt)\right\}\nonumber \\
&&\hspace{-1cm}-2\sum_{j<k}\left|U_{\alpha j}U_{\alpha k}\right|^2
\left\{2\sin^2 \left(\frac{\Delta E_{jk}t}{2}\right)+\frac{E_jE_k-p^2}{E_jE_k}\sin (E_jt)\sin (E_kt)\right\},
\\
P(\nu_{\alpha L}^{c\prime}\to\nu_{\beta L}^{c\prime})&=&
-\sum_j |U_{\alpha j}U_{\beta j}|^2\left\{\frac{m_j^2}{E_j^2}\sin^2 (E_jt)\right\}\nonumber \\
&&\hspace{-1.5cm}-2\sum_{j<k}{\rm Re}\left[U_{\alpha j}U_{\beta j}^*U_{\alpha k}^*U_{\beta k}\right]
\left\{2\sin^2 \left(\frac{\Delta E_{jk}t}{2}\right)+\frac{E_jE_k-p^2}{E_jE_k}\sin (E_jt)\sin (E_kt)\right\} \nonumber \\
&&\hspace{-1.5cm}-2\sum_{j<k}{\rm Im}\left[U_{\alpha j}U_{\beta j}^*U_{\alpha k}^*U_{\beta k}\right]
\left\{\sin (\Delta E_{jk}t)+\frac{E_k-p}{E_k}\cos (E_jt)\sin (E_kt)-\frac{E_j-p}{E_j}\cos (E_kt)\sin (E_jt)\right\}, \\
P(\nu_{\alpha L}^{c\prime}\to\nu_{\alpha L}^{\prime})&=&
\sum_j |U_{\alpha j}U_{\alpha j}|^2\frac{m_j^2}{E_j^2}\sin^2 (E_jt)
+2\sum_{j<k}{\rm Re}[U_{\alpha j}U_{\alpha k}^*U_{\alpha j}U_{\alpha k}^*]\frac{m_jm_k}{E_jE_k}\sin (E_jt)\sin (E_kt),\\
P(\nu_{\alpha L}^{c\prime}\to\nu_{\beta L}^{\prime})&=&
\sum_j |U_{\alpha j}U_{\beta j}|^2\frac{m_j^2}{E_j^2}\sin^2 (E_jt)
+2\sum_{j<k}{\rm Re}[U_{\alpha j}U_{\alpha k}^*U_{\beta j}U_{\beta k}^*]\frac{m_jm_k}{E_jE_k}\sin (E_jt)\sin (E_kt).
\end{eqnarray}
%\end{widetext}
Comparing these probabilities with (\ref{aLp-aLp})-(\ref{aLp-bLcp}),
The only difference from the corresponding neutrino probabilities is the sign of the term
proportional to ${\rm Im}\left[U_{\alpha j}U_{\beta j}^*U_{\alpha k}^*U_{\beta k}\right]$.

%%%%%%%%%%%%%%%%%%%%%%%%%%%%%
\section{Relation of Oscillation Probabilities}

Next, let us investigate the relationship between the CP-conjugate probabilities or the T-conjugate probabilities
both in the case for the Dirac and the Majorana neutrinos.
In order to do that, we summarize the relation of probabilities.
In the Dirac neutrinos, we obtain the following relations about the probabilities without chirality-flip,
%\begin{widetext}
\begin{eqnarray}
&&P(\nu_{\alpha L}\to \nu_{\alpha L})=P(\nu_{\alpha L}^{\prime}\to \nu_{\alpha L}^{\prime})
=P(\nu_{\alpha L}^{c}\to \nu_{\alpha L}^{c})=P(\nu_{\alpha L}^{c\prime}\to \nu_{\alpha L}^{c\prime}), \\
&&P(\nu_{\alpha R}\to \nu_{\alpha R})=P(\nu_{\alpha R}^{\prime}\to \nu_{\alpha R}^{\prime})
=P(\nu_{\alpha R}^{c}\to \nu_{\alpha R}^{c})=P(\nu_{\alpha R}^{c\prime}\to \nu_{\alpha R}^{c\prime}), \\
&&P(\nu_{\alpha L}\to \nu_{\beta L})=P(\nu_{\beta L}^{c}\to \nu_{\alpha L}^{c})
=P(\nu_{\beta L}^{\prime}\to \nu_{\alpha L}^{\prime})=P(\nu_{\alpha L}^{c\prime}\to \nu_{\beta L}^{c\prime}),
\label{im-plus-L} \\
&&P(\nu_{\beta L}\to \nu_{\alpha L})=P(\nu_{\alpha L}^{c}\to \nu_{\beta L}^{c})
=P(\nu_{\alpha L}^{\prime}\to \nu_{\beta L}^{\prime})=P(\nu_{\beta L}^{c\prime}\to \nu_{\alpha L}^{c\prime}),
\label{im-minus-L} \\
&&P(\nu_{\alpha R}\to \nu_{\beta R})=P(\nu_{\beta R}^{c}\to \nu_{\alpha R}^{c})
=P(\nu_{\beta R}^{\prime}\to \nu_{\alpha R}^{\prime})=P(\nu_{\alpha R}^{c\prime}\to \nu_{\beta R}^{c\prime}),
\label{im-plus-R} \\
&&P(\nu_{\beta R}\to \nu_{\alpha R})=P(\nu_{\alpha R}^{c}\to \nu_{\beta R}^{c})
=P(\nu_{\alpha R}^{\prime}\to \nu_{\beta R}^{\prime})=P(\nu_{\beta R}^{c\prime}\to \nu_{\alpha R}^{c\prime}).
\label{im-minus-R}
\end{eqnarray}
The difference of (\ref{im-plus-L}) and (\ref{im-minus-L}) and the difference of
(\ref{im-plus-R}) and (\ref{im-minus-R}) are both sign of the term proportional to
${\rm Im}\left[U_{\alpha j}U_{\beta j}^*U_{\alpha k}^*U_{\beta k}\right]$ and
${\rm Im}\left[V_{\alpha j}V_{\beta j}^*V_{\alpha k}^*V_{\beta k}\right]$ respectively.
The probabilities with chirality-flip have the relations,
\begin{eqnarray}
&&P(\nu_{\alpha L}\to \nu_{\alpha R})=P(\nu_{\alpha L}^{\prime}\to \nu_{\alpha R}^{\prime})
=P(\nu_{\alpha L}^{c}\to \nu_{\alpha R}^{c})=P(\nu_{\alpha L}^{c\prime}\to \nu_{\alpha R}^{c\prime}), \\
&&P(\nu_{\alpha L}\to \nu_{\beta R})=P(\nu_{\alpha L}^{\prime}\to \nu_{\beta R}^{\prime})
=P(\nu_{\alpha L}^{c}\to \nu_{\beta R}^{c})=P(\nu_{\alpha L}^{c\prime}\to \nu_{\beta R}^{c\prime}) \\
&&=P(\nu_{\beta R}\to \nu_{\alpha L})=P(\nu_{\beta R}^{\prime}\to \nu_{\alpha L}^{\prime})
=P(\nu_{\beta R}^{c}\to \nu_{\alpha L}^{c})=P(\nu_{\beta R}^{c\prime}\to \nu_{\alpha L}^{c\prime}).
\end{eqnarray}

Next, the relations on the Majorana neutrino oscillation probabilities without chirality-flip are given by
\begin{eqnarray}
&&P(\nu_{\alpha L}\to \nu_{\alpha L})=P(\nu_{\alpha L}^{\prime}\to \nu_{\alpha L}^{\prime})
=P(\nu_{\alpha L}^{c}\to \nu_{\alpha L}^{c})=P(\nu_{\alpha L}^{c\prime}\to \nu_{\alpha L}^{c\prime}), \\
&&P(\nu_{\alpha L}\to \nu_{\beta L})=P(\nu_{\beta L}^{c}\to \nu_{\alpha L}^{c})
=P(\nu_{\beta L}^{\prime}\to \nu_{\alpha L}^{\prime})=P(\nu_{\alpha L}^{c\prime}\to \nu_{\beta L}^{c\prime}),
\label{im-plus-L-majo} \\
&&P(\nu_{\beta L}\to \nu_{\alpha L})=P(\nu_{\alpha L}^{c}\to \nu_{\beta L}^{c})
=P(\nu_{\alpha L}^{\prime}\to \nu_{\beta L}^{\prime})=P(\nu_{\beta L}^{c\prime}\to \nu_{\alpha L}^{c\prime}).
\label{im-minus-L-majo}
\end{eqnarray}
The difference between (\ref{im-plus-L-majo}) and (\ref{im-minus-L-majo}) is also the sign of the term
proportional to ${\rm Im}\left[U_{\alpha j}U_{\beta j}^*U_{\alpha k}^*U_{\beta k}\right]$.
About the oscillation probabilities with chirality-flip, we have the following relations,
\begin{eqnarray}
&&P(\nu_{\alpha L}\to \nu_{\alpha L}^c)=P(\nu_{\alpha L}^{\prime}\to \nu_{\alpha L}^{c\prime})
=P(\nu_{\alpha L}^{c}\to \nu_{\alpha L})=P(\nu_{\alpha L}^{c\prime}\to \nu_{\alpha L}^{\prime}), \\
&&P(\nu_{\alpha L}\to \nu_{\beta L}^c)=P(\nu_{\alpha L}^{\prime}\to \nu_{\beta L}^{c\prime})
=P(\nu_{\alpha L}^{c}\to \nu_{\beta L})=P(\nu_{\alpha L}^{c\prime}\to \nu_{\beta L}^{\prime}) \\
&&=P(\nu_{\beta L}^c\to \nu_{\alpha L})=P(\nu_{\beta L}^{c\prime}\to \nu_{\alpha L}^{\prime})
=P(\nu_{\beta L}\to \nu_{\alpha L}^{c})=P(\nu_{\beta L}^{\prime}\to \nu_{\alpha L}^{c\prime}).
\end{eqnarray}
%\end{widetext}

Next, let us present the differences between CP-conjugate probabilities,
T-conjugate probabilities and CPT-conjugate probabilities.
They are obtained by the following replacement
in an original probability,
\begin{eqnarray}
&&\hspace{-0.5cm}{\rm CP \,conjugate}: U \leftrightarrow U^*, V \leftrightarrow V^*, \nu \leftrightarrow \nu^{c},
\nonumber \\
&&\hspace{-0.5cm}{\rm T \,conjugate} : \alpha \leftrightarrow \beta, \nonumber \\
&&\hspace{-0.5cm}{\rm CPT \,conjugate}: U \leftrightarrow U^*, V \leftrightarrow V^*, \nu \leftrightarrow \nu^{c},
\alpha \leftrightarrow \beta
\nonumber \\
\end{eqnarray}

First, in the case of the Dirac neutrinos, they are respectively given by
%\begin{widetext}
\begin{eqnarray}
&&\Delta P_{\rm CP}(\nu_{\alpha L}\to\nu_{\beta L})=P(\nu_{\alpha L}\to\nu_{\beta L})-P(\nu_{\alpha L}^{c}\to\nu_{\beta L}^{c})\nonumber \\
&&\hspace{0.7cm}=-4\sum_{j<k}{\rm Im}\left[U_{\alpha j}U_{\beta j}^*U_{\alpha k}^*U_{\beta k}\right]
\left\{\sin (\Delta E_{jk}t)+\frac{E_k-p}{E_k}\cos (E_jt)\sin (E_kt)-\frac{E_j-p}{E_j}\cos (E_kt)\sin (E_jt)\right\},\\
&&\Delta P_{\rm CP}(\nu_{\alpha R}\to\nu_{\beta R})=P(\nu_{\alpha R}\to\nu_{\beta R})-P(\nu_{\alpha R}^{c}\to\nu_{\beta R}^{c})\nonumber \\
&&\hspace{0.7cm}=4\sum_{j<k}{\rm Im}\left[V_{\alpha j}V_{\beta j}^*V_{\alpha k}^*V_{\beta k}\right]
\left\{\sin (\Delta E_{jk}t)+\frac{E_k-p}{E_k}\cos (E_jt)\sin (E_kt)-\frac{E_j-p}{E_j}\cos (E_kt)\sin (E_jt)\right\},\\
&&\Delta P_{\rm CP}(\nu_{\alpha L}\to\nu_{\beta R})=P(\nu_{\alpha L}\to\nu_{\beta R})-P(\nu_{\alpha L}^{c}\to\nu_{\beta R}^{c})=0,\\
&&\Delta P_{\rm T}(\nu_{\alpha L}\to\nu_{\beta L})=P(\nu_{\alpha L}\to\nu_{\beta L})-P(\nu_{\beta L}\to\nu_{\alpha L})
\nonumber \\
&&\hspace{0.7cm}=-4\sum_{j<k}{\rm Im}\left[U_{\alpha j}U_{\beta j}^*U_{\alpha k}^*U_{\beta k}\right]
\left\{\sin (\Delta E_{jk}t)+\frac{E_k-p}{E_k}\cos (E_jt)\sin (E_kt)-\frac{E_j-p}{E_j}\cos (E_kt)\sin (E_jt)\right\},\\
&&\Delta P_{\rm T}(\nu_{\alpha R}\to\nu_{\beta R})=P(\nu_{\alpha R}\to\nu_{\beta R})-P(\nu_{\beta R}\to\nu_{\alpha R})
\nonumber \\
&&\hspace{0.7cm}=4\sum_{j<k}{\rm Im}\left[V_{\alpha j}V_{\beta j}^*V_{\alpha k}^*V_{\beta k}\right]
\left\{\sin (\Delta E_{jk}t)+\frac{E_k-p}{E_k}\cos (E_jt)\sin (E_kt)-\frac{E_j-p}{E_j}\cos (E_kt)\sin (E_jt)\right\},\\
&&\Delta P_{\rm T}(\nu_{\alpha L}\to\nu_{\beta R})=P(\nu_{\alpha L}\to\nu_{\beta R})-P(\nu_{\beta R}\to\nu_{\alpha L})=0, \\
&&\Delta P_{\rm CPT}(\nu_{\alpha L}\to\nu_{\beta L})
=P(\nu_{\alpha L}\to\nu_{\beta L})-P(\nu_{\beta L}^{c}\to\nu_{\alpha L}^{c})=0, \\
&&\Delta P_{\rm CPT}(\nu_{\alpha R}\to\nu_{\beta R})
=P(\nu_{\alpha R}\to\nu_{\beta R})-P(\nu_{\beta R}^{c}\to\nu_{\alpha R}^{c})=0, \\
&&\Delta P_{\rm CPT}(\nu_{\alpha L}\to\nu_{\beta R})
=P(\nu_{\alpha L}\to\nu_{\beta R})-P(\nu_{\beta R}^{c}\to\nu_{\alpha L}^{c})=0.
\end{eqnarray}
Second, in the case of the Majorana neutrinos, they are respectively given by
\begin{eqnarray}
&&\Delta P_{\rm CP}(\nu_{\alpha L}\to\nu_{\beta L})=P(\nu_{\alpha L}\to\nu_{\beta L})-P(\nu_{\alpha L}^{c}\to\nu_{\beta L}^{c})\nonumber \\
&&\hspace{0.7cm}=-4\sum_{j<k}{\rm Im}\left[U_{\alpha j}U_{\beta j}^*U_{\alpha k}^*U_{\beta k}\right]
\left\{\sin (\Delta E_{jk}t)+\frac{E_k-p}{E_k}\cos (E_jt)\sin (E_kt)-\frac{E_j-p}{E_j}\cos (E_kt)\sin (E_jt)\right\},\\
&&\Delta P_{\rm CP}(\nu_{\alpha L}\to\nu_{\beta L}^c)=P(\nu_{\alpha L}\to\nu_{\beta L}^c)-P(\nu_{\alpha L}^{c}\to\nu_{\beta L})=0,\\
&&\Delta P_{\rm T}(\nu_{\alpha L}\to\nu_{\beta L})=P(\nu_{\alpha L}\to\nu_{\beta L})-P(\nu_{\beta L}\to\nu_{\alpha L})
\nonumber \\
&&\hspace{0.7cm}=-4\sum_{j<k}{\rm Im}\left[U_{\alpha j}U_{\beta j}^*U_{\alpha k}^*U_{\beta k}\right]
\left\{\sin (\Delta E_{jk}t)+\frac{E_k-p}{E_k}\cos (E_jt)\sin (E_kt)-\frac{E_j-p}{E_j}\cos (E_kt)\sin (E_jt)\right\},\\
&&\Delta P_{\rm T}(\nu_{\alpha L}\to\nu_{\beta L}^c)=P(\nu_{\alpha L}\to\nu_{\beta L}^c)-P(\nu_{\beta L}^c\to\nu_{\alpha L})=0, \\
&&\Delta P_{\rm CPT}(\nu_{\alpha L}\to\nu_{\beta L})
=P(\nu_{\alpha L}\to\nu_{\beta L})-P(\nu_{\beta L}^{c}\to\nu_{\alpha L}^{c})=0, \\
&&\Delta P_{\rm CPT}(\nu_{\alpha L}\to\nu_{\beta L}^c)
=P(\nu_{\alpha L}\to\nu_{\beta L}^c)-P(\nu_{\beta L}\to\nu_{\alpha L}^{c})=0.
\end{eqnarray}
\end{widetext}
These results hold even when we extend to n-generations.
In the case of both Dirac and Majorana neutrinos, there is no direct CP and T violation related to the new CP phases
in vacuum.
Namely, the differences between the original and the CP or T-conjugate probabilities vanish in the oscillations with chirality-flip.
This means that we cannot explain the reason for the existence of matter in the universe by neutrino oscillations
in vacuum even if neutrinos are the Majorana particles.

%%%%%%%%%%%%%%%%%%%%%%%%%%%%%%%%%%%%%%%%%%%%%%%%%
\section{Comparison of Conventional Result and New Result for Majorana Neutrinos}

In this section, we review the previous results \cite{Bahcall1978, Valle1981, Li1982, Bernabeu1983, Gouvea2003, Xing2013} 
on the probabilities for $\nu \leftrightarrow \nu^c$ oscillations
and compare with our results.
The amplitudes in previous papers are given by
\begin{eqnarray}
A(\nu_{\alpha L}\to \nu_{\beta L}^{c})&=&\sum_j \left[U_{\alpha j}^*U_{\beta j}^*\frac{m_j}{E_j}e^{-iE_jt}\right]K, \\
A(\nu_{\alpha L}^{c}\to \nu_{\beta L})&=&\sum_j \left[U_{\alpha j}U_{\beta j}\frac{m_j}{E_j}e^{-iE_jt}\right]\bar{K},
\end{eqnarray}
where $K$ and $\bar{K}$ are the kinematical factors independent of the index $j$ (and sastisfying $|K|=|\bar{K}|$).
On the other hand, our results are
\begin{widetext}
\begin{eqnarray}
A(\nu_{\alpha L}\to\nu_{\beta L}^{c})&=&\sum_j U_{\alpha j}^*U_{\beta j}^*\frac{m_j}{E_j}(e^{-iE_jt}-e^{iE_jt})
=-i\sum_j U_{\alpha j}^*U_{\beta j}^*\frac{m_j}{E_j}\sin (E_jt), \\
A(\nu_{\alpha L}^{c}\to\nu_{\beta L})&=&\sum_j U_{\alpha j}U_{\beta j}\frac{m_j}{E_j}(e^{iE_jt}-e^{-iE_jt})
=i\sum_j U_{\alpha j}U_{\beta j}\frac{m_j}{E_j}\sin (E_jt).
\end{eqnarray}
The difference of our result from the previous result is in the negative energy part proportional to $e^{iE_jt}$.
In the case that we calculate the oscillation probabilities based on the Dirac equation,
$\nu$ and $\nu^c$ are included in the same multiplet and a state of the neutrino is represented as the linear
combination of both positive and negative energy parts.

Next, we compare the oscillation probabilities.
The probabilities presented in the previous papers are given by
%\begin{widetext}
\begin{eqnarray}
P(\nu_{\alpha L}\to \nu_{\beta L}^{c})&=&\frac{|K|^2}{E^2}\left[\sum_j \left|m_j U_{\alpha j}^*U_{\beta j}^*\right|^2+
2\sum_{j<k}m_jm_k {\rm Re}\left(U_{\alpha j}^*U_{\beta j}^*U_{\alpha k}U_{\beta k}e^{-i\Delta E_{jk}t}\right)\right]
\nonumber \\
&&\hspace{-3cm}=\frac{|K|^2}{E^2}\left[\sum_j \left|m_j U_{\alpha j}U_{\beta j}\right|^2+
2\sum_{j<k}m_jm_k \left\{{\rm Re}\left(U_{\alpha j}U_{\beta j}U_{\alpha k}^*U_{\beta k}^*\right)\cos(\Delta E_{jk}t)
-{\rm Im}\left(U_{\alpha j}U_{\beta j}U_{\alpha k}^*U_{\beta k}^*\right)\sin(\Delta E_{jk}t)\right\}\right] \nonumber \\
&&\hspace{-3cm}=\frac{|K|^2}{E^2}\left[\left|\langle m\rangle_{\alpha\beta}\right|^2-
2\sum_{j<k}m_jm_k \left\{2{\rm Re}\left(U_{\alpha j}U_{\beta j}U_{\alpha k}^*U_{\beta k}^*\right)
\sin^2 \left(\frac{\Delta E_{jk}t}{2}\right)
+{\rm Im}\left(U_{\alpha j}U_{\beta j}U_{\alpha k}^*U_{\beta k}^*\right)\sin(\Delta E_{jk}t)\right\}\right],
\label{p-previous}\\
P(\nu_{\alpha L}^{c}\to \nu_{\beta L})&=&\frac{K^2}{E^2}\left[\sum_j \left|m_j U_{\alpha j}U_{\beta j}\right|^2+
2\sum_{j<k}m_jm_k {\rm Re}\left(U_{\alpha j}U_{\beta j}U_{\alpha k}^*U_{\beta k}^*e^{-i\Delta E_{jk}t}\right)\right]
\nonumber \\
&&\hspace{-3cm}=\frac{|\bar{K}|^2}{E^2}\left[\left|\langle m\rangle_{\alpha\beta}\right|^2-
2\sum_{j<k}m_jm_k \left\{2{\rm Re}\left(U_{\alpha j}U_{\beta j}U_{\alpha k}^*U_{\beta k}^*\right)
\sin^2 \left(\frac{\Delta E_{jk}t}{2}\right)
-{\rm Im}\left(U_{\alpha j}U_{\beta j}U_{\alpha k}^*U_{\beta k}^*\right)\sin(\Delta E_{jk}t)\right\}\right],
\end{eqnarray}
where
\begin{eqnarray}
\left|\langle m\rangle_{\alpha\beta}\right|\equiv \left|\sum_j m_j U_{\alpha j}U_{\beta j}\right|
\end{eqnarray}
is effective mass of the Majorana neutrinos.
Accordingly, there is a difference between CP-conjugate probailities,
\begin{eqnarray}
P(\nu_{\alpha L}\to\nu_{\beta L}^{c})-P(\nu_{\alpha L}^{c}\to\nu_{\beta L})
%\nonumber \\
=\frac{|K|^2}{E^2}\left[-4\sum_{j<k}m_jm_k {\rm Im}\left(U_{\alpha j}U_{\beta j}U_{\alpha k}^*U_{\beta k}^*\right)\sin(\Delta E_{jk}t)\right].
\end{eqnarray}
%\end{widetext}

On the contrary, in this paper, we have the same probability for CP-conjugate probabilities as
\begin{eqnarray}
P(\nu_{\alpha L}\to\nu_{\beta L}^{c})=P(\nu_{\alpha L}^{c}\to\nu_{\beta L})
=\left|\sum_j \frac{m_j}{E_j}U_{\alpha j}U_{\beta j}\sin (E_jt)\right|^2
\simeq \frac{1}{2E^2}\left|\langle m\rangle_{\alpha\beta}\right|^2, \label{p-our}
\end{eqnarray}
\end{widetext}
where the last term is obtained by the averaging the sine term.
Then, the difference between CP-conjugate probabilities becomes
\begin{eqnarray}
P(\nu_{\alpha L}\to\nu_{\beta L}^{c})-P(\nu_{\alpha L}^{c}\to\nu_{\beta L})=0.
\end{eqnarray}
Therefore, we found that the CP violation due to the Majorana CP phase does not appear
even if we consider $\nu \leftrightarrow \nu^c$ oscillations with different flavor.

Another difference is the probability at zero-distance.
In the previous papers \cite{Li1982}, it has been pointed out the zero-distance effect. Namely, if we take the limit of $t \to 0$
in eq.(\ref{p-previous}), the probability has non-zero value,
\begin{eqnarray}
P(\nu_{\alpha L}\to \nu_{\beta L}^{c})=\frac{|K|^2}{E^2}\left|\langle m\rangle_{\alpha\beta}\right|^2.
\end{eqnarray}
However, there is no zero-distance effect in our result from eq.(\ref{p-our}).

\section{Summary}

In three generations, we have derived the exact neutrino oscillation probabilities relativistically
by using the Dirac equation. The results obtained in the three generations can be extended to the case of
$n$ generations.
We have calculated various oscillation probabilities both in the Dirac neutrinos and the Majorana neutrinos.
These probabilities can be calculated by the same formulation and can be understood in a unified way.
The oscillation probabilities about the Dirac neutrinos derived in this paper are classified as
\begin{itemize}
\item the probabilities from left-handed neutrino with negative helicity $\nu_{\alpha L}$ to other neutrinos
\item the probabilities from left-handed neutrino with positive helicity $\nu_{\alpha L}^{\prime}$ to other neutrinos
\item the probabilities from right-handed neutrino with positive helicity $\nu_{\alpha R}$ to other neutrinos
\item the probabilities from right-handed neutrino with negative helicity $\nu_{\alpha R}^{\prime}$ to other neutrinos
\item the probabilities from right-handed anti-neutrino with positive helicity $\nu_{\alpha L}^c$ to other neutrinos
\item the probabilities from right-handed anti-neutrino with negative helicity $\nu_{\alpha L}^{c\prime}$ to other neutrinos
\item the probabilities from left-handed anti-neutrino with negative helicity $\nu_{\alpha R}^c$ to other neutrinos
\item the probabilities from left-handed anti-neutrino with positive helicity $\nu_{\alpha R}^{c\prime}$ to other neutrinos
\end{itemize}
In these probabilities, both oscillations with and without chirality-flip are included.
About the Majorana neutrinos, the probabilities are classified as
\begin{itemize}
\item the probabilities from left-handed neutrino with negative helicity $\nu_{\alpha L}$ to other neutrinos
\item the probabilities from left-handed neutrino with positive helicity $\nu_{\alpha L}^{\prime}$ to other neutrinos
\item the probabilities from right-handed anti-neutrino with positive helicity $\nu_{\alpha L}^c$ to other neutrinos
\item the probabilities from right-handed anti-neutrino with negative helicity $\nu_{\alpha L}^{c\prime}$ to other neutrinos
\end{itemize}
In these probabilities, the oscillations between neutrinos and anti-neutrinos are included.
These probabilities are not independent but related to each other.

As neutrinos have finite mass, there are two components for each chirality
corresponding to positive and negative helicities.
We have shown that the probability is different for each component even if neutrinos have the same chirality.
We have also shown the probabilities depend on not only the mass squared differences but also the absolute masses of neutrinos.
Besides, the new CP phases appear in the probabilities of oscillations with chirality-flip.
These new CP phases are equivalent to the Majorana CP phases in the case of Majorana neutrinos.
We have also investigated the CP dependence of oscillation probabilities in vacuum
and counted the number of the CP phases in $n$ generations.

In the case of Majorana neutrinos, there is no direct CP violation in $\nu_{\alpha}\leftrightarrow \nu_{\beta}^c$ oscillations even if the flavors, $\alpha$ and $\beta$, are different as in the same as two generations \cite{KT2}.
In other words, the difference between the CP-conjugate probabilities 
$P(\nu_{\alpha L}\to \nu_{\beta L}^c)-P(\nu_{\alpha L}^c\to \nu_{\beta L})$ vanishes.
Although there is only indirect CP violation, we obtain the information of the new CP phases
through both cosine and sine terms.
So, we can determine the value of the CP phases.
Furthermore, it has been said that the zero-distance effect appears in the oscillations
between neutrinos and anti-neutrinos with different flavors in the Majorana neutrino case.
However, we have shown that the zero-distance effect does not appear in our formulation.
These are different from the results written in previous papers \cite{Bahcall1978, Valle1981, Li1982, Bernabeu1983, Gouvea2003, Xing2013}.

\end{document}